\documentclass[fleqn,usenatbib]{mnras} %a4paper

\usepackage{graphicx}	% Including figure files
\usepackage{amsmath}	% Advanced maths commands
\usepackage{amssymb}	% Extra maths symbols

\usepackage{multirow}

\newcommand{\kms}{km\,s$^{-1}$} % km per second

\title[The P-q relation of wide sdB+MS binaries]{The orbital period -- mass ratio relation of wide sdB+MS binaries and its application to the stability of RLOF.}

\author[J. Vos et al.]{
Joris Vos$^{1,2}$\thanks{E-mail: astro@jorisvos.eu},
Maja Vu\u{c}kovi\'{c}$^{1}$,
Xuefei Chen$^{3,4,5}$,
Zhanwen Han$^{3,4,5}$,
Thomas Boudreaux$^{6}$,
\newauthor
Brad N. Barlow$^{6}$,
Roy \O{}stensen$^{7}$,
P\'{e}ter N\'{e}meth$^{8, 9}$
\\
$^{1}$Instituto de F\'{\i}sica y Astronom\'{\i}a, Universidad de Valparaiso, Gran Breta\~{n}a 1111, Playa Ancha, Valpara\'{\i}so
2360102, Chile\\
$^{2}$Institut f\"{u}r Physik und Astronomie, Universit\"{a}t Potsdam, Karl-Liebknecht-Str. 24/25, 14476, Golm, Germany\\
$^{3}$Yunnan Observatories, Chinese Academy of Sciences, 396 Yangfangwang, Guandu District, Kunming, 650216, P. R. China\\
$^{4}$Key Laboratory for the Structure and Evolution of Celestial Objects, Chinese Academy of Sciences, 396 Yangfangwang, \\Guandu District, Kunming, 650216, P. R. China\\
$^{5}$Center for Astronomical Mega-Science, Chinese Academy of Sciences, 20A Datun Road, Chaoyang District, Beijing, 100012, P. R. China\\
$^{6}$Department of Physics, High Point University, One University Parkway, High Point, NC 27268, USA\\
$^{7}$Department of Physics, Astronomy, and Materials Science, Missouri State University, Springfield, MO 65804, USA\\
$^{8}$Astronomical Institute of the Czech Academy of Sciences, CZ-251\,65, Ond\v{r}ejov, Czech Republic\\
$^{9}$Astroserver.org, 8533 Malomsok, Hungary\\
}

\date{Accepted XXX. Received YYY; in original form ZZZ}

\pubyear{2018}

\begin{document}
\label{firstpage}
\pagerange{\pageref{firstpage}--\pageref{lastpage}}
\maketitle

\begin{abstract}
Wide binaries with hot subdwarf-B (sdB) primaries and main sequence companions are thought to form only through stable Roche lobe overflow (RLOF) of the sdB progenitor near the tip of the red giant branch (RGB). We present the orbital parameters of eleven new long period composite sdB binaries based on spectroscopic observations obtained with the UVES, FEROS and CHIRON spectrographs. Using all wide sdB binaries with known orbital parameters, 23 systems, the observed period distribution is found to match very well with theoretical predictions. A second result is the strong correlation between the orbital period (P) and the mass ratio (q) in the observed wide sdB binaries. In the P-q plane two distinct groups emerge, with the main group (18 systems) showing a strong correlation of lower mass ratios at longer orbital periods. The second group are systems that are thought to be formed from higher mass progenitors. Based on theoretical models, a correlation between the initial mass ratio at the start of RLOF and core mass of the sdB progenitor is found, which defines a mass-ratio range at which RLOF is stable on the RGB.
\end{abstract}


%%%%%%%%%%%%%%%%%%%%%%%%%%%%%%%%%%%%%%%%%%%%%%%%%%

%%%%%%%%%%%%%%%%% BODY OF PAPER %%%%%%%%%%%%%%%%%%

\section{Introduction}
Hot subdwarf-B (sdB) stars are core-helium-burning stars with a very thin hydrogen envelope (M$_{\rm{H}}$ $<$ 0.02 $M_{\odot}$), and a mass close to the core-helium-flash mass $\sim$ 0.47 $M_{\odot}$ \citep{Heber2009, Heber2016}. It was found that many sdB stars reside in a binary system \citep[e.g.][]{Koen1998, Maxted2001, Morales2003, Napiwotzki2004}. Currently the consensus is that sdBs are solely formed by binary interaction, and the three main formation channels that contribute to the sdB population are the common-envelope (CE) ejection channel \citep{Paczynski1976, Han2002}, the stable Roche-lobe overflow (RLOF) channel \citep{Han2000, Han2002}, and the formation of a single sdB star as the end product of a binary white-dwarf (WD) merger \citep{Webbink1984}.

Binary population synthesis (BPS) studies performed by \citet{Han2002, Han2003} and \citet{Chen2013} found that the CE ejection channel leads to close binaries with periods on the order of hours up to tens of days with white dwarf (WD) or main sequence (MS) companions. Many observational studies have focused on these short period systems \citep[e.g.][]{Copperwheat2011, Geier2011, Kupfer2015}, and the more than 150 solved systems match well with the results from BPS studies. Furthermore, also possible progenitors for the close sdB binaries have been proposed \citep[see e.g.][]{Beck2014} The RLOF channel produces sdB+MS binaries with orbital periods on the order of several years, up to $\sim$ 1600 days. Long period sdB binaries have been sugested \citep[e.g.][]{Green2001}, but only recently the first orbital periods have been published \citep{Deca2012, Oestensen2012}. Finaly, the WD merger channel produces single sdBs which can potentially have higher masses. Here we focus on the wide sdB+MS binaries formed through the stable RLOF channel. 

One of the main unknowns in the evolution of binary systems containing a red giant star is the stability of RLOF. This stability criterion has an important impact on the final orbital periods of those systems that undergo mass loss on the RGB. If the sdB-progenitor's evolutionary expansion on the RGB matches that of its Roche-lobe, mass loss will be stable. However, if the increase in radius of the primary significantly exceeds that of its Roche-lobe, a CE will be formed and the system will undergo a spiral-in phase. Currently employed stability criteria based on polytropic models \citep{Hjellming1987} are clearly too strict \citep{Woods2012}. Improvements on these criteria have been suggested by \citet{Ge2015} using adiabatic models, and \citet{Pavlovskii2015} found that a super-adiabatic layer in the donor star allows for stable mass-loss in a wider range of conditions.

Wide sdB+MS binaries are useful systems to study the stability of RLOF on the RGB. They are double lined spectroscopic binaries of which the primary mass is close to the core-helium-flash mass, allowing a complete solution for the orbits. A long term observing program was started in 2009 with the HERMES spectrograph at the Mercator telescope \citep{Oestensen2012}. This program was extended in 2011 to cover southern targets using UVES at the 8.2m VLT, FEROS at the 2.2m MPG and CHIRON at the 1.5m SMARTS telescope \citep{Vos2018a}. Currently 36 wide sdB binaries are being monitored in this program.
The systems that were part of the original program with Mercator have been analysed and orbital parameters for eight systems have been published in \citet{Vos2012, Vos2013, Vos2017}. Here we present the orbital parameters of eleven new systems observed with the UVES, FEROS and CHIRON spectrographs. These eleven new systems allow us to study the relation between orbital period and mass-ratio for all known long period sdB+MS binaries with solved orbits. The focus of this article is on the orbital parameters of the wide sdB+MS binaries. A detailed study of their atmospheric properties and population membership will be the subject of a future article. 

Using all wide sdB+MS binaries with known orbital periods, we compare the distribution of the observed periods with the predictions of the theoretical models of \cite{Chen2013}. Furthermore the relation between the mass ratio and orbital period is used to study the stability of RLOF on the RGB, and derive a relation between the critical initial mass ratio at the start of RLOF and the final mass of the hot subdwarf star. 

\section{Spectroscopy}\label{s:spectroscopy}
The wide sdB+MS systems presented in this article are part of a long term monitoring campaign using the UVES spectrograph at the VLT Kueyen telescope (UT2) on Cerro Paranal described in \citet{Vos2018a}. UVES is a two-arm cross-dispersed echelle spectrograph \citep{Dekker2000}, and it was used in standard dichroic-2 437+760 mode covering a wavelength range of 3730 - 4990 \AA\ in the BLUE arm and 5650 - 9460 \AA\ in the RED arm. To reach a resolution of around 40\,000, a slit width of 1 arcsec was used. At this moment ten of these systems have sufficient orbital coverage to derive orbital parameters. An overview of these systems is given in Table\,\ref{tb:observations}. All UVES spectra were reduced using the UVES pipeline and the {\sc reflex} workflow engine \citep{reflex2013}.

For five of these targets earlier observations taken with FEROS at the 2.2m MPG telescope are available in the ESO archives. These observations are ideal to constrain the orbital period. To reduce the FEROS spectra, the Collection of Elemental Routines for Echelle Spectra \citep[CERES,][]{Brahm2017} package was used. CERES is a set of routines for echelle spectrographs, which contains a fully automated pipeline to reduce FEROS spectra. 

One system was observed with the fiber fed echelle spectrograph CHIRON \citep{Tokovinin2013} attached to the 1.5m telescope at the Cerro Tololo Inter-American Observatory in Chile operated by the SMARTS consortium. CHIRON was used in fiber mode with an average resolution of R = 25\,000, and the spectra have an average signal to noise (S/N) of 30. The CHIRON spectra cover the wavelength range from 4505 to 8899 \AA\ in 62 orders. To increase the wavelength stability a ThAr spectrum was taken after each observation. The spectra were reduced and wavelength calibrated by the SMARTS consortium using the Yale pipeline \citep{Tokovinin2013}.

\begin{table*}
   \centering
   \caption{The coordinates and magnitudes and classification of the eleven new wide sdB binaries. The classification provided here is taken from the literature. The spectrographs used are UVES at the VLT (U), FEROS at the 2.2m MPG (F) and CHIRON at the 1.5m SMARTS (C). In the last column is indicated if the sdB component has metal lines visible in the spectrum which were used in the determinaton of their radial velocities.} \label{tb:observations}
   \begin{tabular}{llllrcc}
    \hline
    Object   &  Class  &  V-mag   &   RA     &   \multicolumn{1}{l}{Dec}      &  Spectrographs  & sdB Metal lines  \\
             &         &          &  (hours) & \multicolumn{1}{l}{(degrees)}  &                 &                  \\\hline \hline

    PB\,6355                   &  sdB+F    &  13.00  &  01 16 27.3  &   +06 03 11.6  &  U/F     &   +   \\          %
    MCT\,0146--2651            &  sdB+F/G  &  12.31  &  01 48 44.0  &  --26 36 12.8  &  U/F/C   &   --   \\
    FAUST\,321                 &  sdB+F    &  13.22  &  01 51 23.4  &  --75 48 38.9  &  U       &   --   \\          %
    JL\,277                    &  sdB+F5   &  13.16  &  02 01 34.4  &  --53 43 43.5  &  U       &   +   \\          %
    GALEX\,J022836.7--362543   &  sdB+K0   &  13.03  &  02 28 36.9  &  --36 25 45.7  &  U       &   +   \\          %
    EC\,03143--5945            &  sdB+F9   &  13.47  &  03 15 30.1  &  --59 34 04.9  &  U/F     &   +   \\          %
    GALEX\,J033216.7--023302   &  sdB+F    &  13.00  &  03 32 16.7  &  --02 33 01.9  &  U/F     &   --   \\          %
    GALEX\,J053939.1--283329   &  sdB+G    &  13.73  &  05 39 39.2  &  --28 33 30.6  &  U       &   --   \\          %
    PG\,1514+034               &  sdOB+G6  &  13.76  &  15 17 14.3  &   +03 10 27.9  &  U       &   --   \\          %
    GALEX\,J162842.0+111838    &  sdB+F    &  13.10  &  16 28 42.0  &   +11 18 39.9  &  U/F     &   --   \\          %
    PG\,2148+095               &  sdB+F    &  13.04  &  21 51 16.9  &   +09 46 59.5  &  U/F     &   +   \\\hline    %
    
   \end{tabular}
\end{table*}

\subsection{Radial velocities}

The determination of the radial velocities of the main sequence (MS) component in the UVES spectra is straightforward as they have many clear metal lines visible in the spectra. Even with a low S/N, accurate velocities can be derived. To derive the radial velocities of the MS component a cross-correlation (CC) with a template spectrum is used. If a spectral analysis of the system has been performed \citep[e.g. in][]{Vos2018a}, a synthetic template with those properties is used. If no such analysis exists, a template matching the spectral class is used. The exact template used for each system is given in the Class information in Table\,\ref{tb:observations}. The CC is performed on regions of both the blue and red orders where there are no significant lines from the sdB star, and where there are no atmospheric or interstellar features. Balmer lines are ignored as well. To derive the radial velocities from the CC function, it is fitted with a Gaussian or rotational template depending on the rotational velocity of the star. To determine the errors of the radial velocities a Monte-Carlo (MC) simulation is used where noise is added to the spectrum and the CCF is repeated, the final error is the standard deviation of the radial velocities determined in this way. This method is explained in more detail in \citet{Vos2017}. The error due to the wavelength stability of the calibrations is also taken into account in this process.

The CERES pipeline used to reduce the FEROS spectra is equipped with a set of functions that can compute the cross correlation function using a binary mask based on the method outlined in \citet{Baranne1996}. The masks used by CERES are the same ones as those used in the HARPS pipeline \citep{Mayor2003}. The CERES pipeline comes with a G2, K5 and M5 mask. As some of the cool companions of our systems are F-type stars, we have supplemented this set with the F0 mask from HARPS. The exact mask used for each system is the one closest to the class given in Table\,\ref{tb:observations}. This algorithm is used to derive the radial velocities of the MS component of our sdB binaries. The only change made to the algorithm is to exclude the bluest orders as the contribution of the MS component in those orders is too small. The errors on the RVs are scaled formal errors where the scaling factor is derived from MC simulations and depends on the S/N of the spectrum and the rotational velocity of the companion \citep{Brahm2017}.

To avoid an offset between the radial velocities of the cool companion derived from the UVES and the FEROS spectra using respectively a template spectrum and a binary mask, we calculate for each system where spectra of both spectrographs are available, the radial velocity offset between the template and the mask. The radial velocities obtained from the FEROS spectra are then corrected for this offset. In all cases the offset is small, with the worst case being on the order of 100 m s$^{-1}$.

The radial velocities of the sdB components are more challenging to determine. In half of the analyzed systems only the \ion{He}{i} blend at 5875.61 \AA\ can be used as other visible lines are contaminated by metal lines from the cool companion. In the other systems the presence of sharp metal lines from C, N and O improve the precision of the RV determination. In Table\,\ref{tb:observations} the last column indiactes with `+' if the sdB component has strong metal lines that could be used in the cross correlation. To derive the radial velocities, a CC with a template spectrum calculated with {\sc Tlusty} \citep{Hubeny1995} is performed. If the spectral parameters of the sdB are known, a template with those parameters is used. Otherwise the best fitting template from a catalog with templates of different effective temperatures and surface gravities is used. The errors on the RVs are determined by using a the same MC method as for the MS companions described earlier. 

The FEROS spectra have in general a lower S/N than the UVES spectra and FEROS is less sensitive in the blue. Therefore the metal lines that are visible in some of the sdBs can not be used in the cross correlation with the FEROS spectra. In a few cases also the \ion{He}{i} blend at 5875.61 \AA\ is too faint to be used and only radial velocities of the MS component could be derived. This is partially caused by the \ion{He}{i} 5875 \AA\ blend falling at the edge of the order and thus has a lower S/N.  As the same template is used to derive the sdBs radial velocities in both spectra, there is no need to correct for an offset between both types of spectra.

To derive the radial velocities from the CHIRON spectra the exact same method as for the FEROS spectra was used, also using the functions that are part of the CERES pipeline. The CHIRON spectra have the advantage that the \ion{He}{i} blend at 5875.61 \AA\ is not lost in the edge of an order, and can be used to derive radial velocities.

The RVs for both the MS and sdB components are given in the appendis in  Tables \ref{tb:rv_PB6355} to \ref{tb:rv_PG2148+095}. Furthermore, they are plotted in Figs.\,\ref{fig:rv_PB6355} to \ref{fig:rv_PG2148+095}.

\subsection{Orbital parameters}

To derive the orbital parameters from the RV curves the same method is followed as described in \citet{Vos2012} and \citet{Vos2013}. A Kepler orbit with eight free parameters, the orbital period ($P$), time of periastron ($T_0$), eccentricity ($e$), angle of periastron ($\omega$), and for both components their radial velocity amplitudes ($K_{\rm{MS}}$ and $K_{\rm{sdB}}$) and systemic velocities ($\gamma_{\rm{MS}}$ and $\gamma_{\rm{sdB}}$), is fitted to the radial velocities. The mass ratio $q$ is defined as $M_{\rm sdB} / M_{\rm MS}$, and is derived from the radial velocity amplitudes $K_{\rm{MS}}$ and $K_{\rm{sdB}}$. The system velocities of both components are fitted independently as the gravitational redshift can cause a significant difference in system velocity between the sdB and the MS component (see e.g. \citealt{Vos2012}). Different from the method used in earlier papers, we employ a Markov-chain-Monte-Carlo (MCMC) approach to determine the errors on the orbital parameters, and check that only one solution is possible. The MCMC method is implemented in {\sc Python} using the affine invariant MCMC ensemble sampler {\sc emcee} of \citet{Foreman-Mackey2013}. 250 walkers are randomly initialized to cover the entire parameters space, including the orbital period which is varied between 300 and 2000 days. Each walker is allowed 1500 steps of which the first 250 are removed to let the walker settle on its fit (burn in). The remaining steps are used to calculate the posterior distribution. For all systems included in this article, only one orbital period emerged from the MCMC fit.

To test if the orbit is significantly eccentric, the \citet{Lucy1971} eccentricity test is used. This test calculates the probability P$_{\rm c}$ of falsely rejecting that the orbit is circular. For low values of P$_{\rm c}$ it would be unreasonable to assume a circular orbit, while for high values the orbit is unlikely to be significantly eccentric. Here we follow the proposal of \citet{Lucy1971} to only accept an eccentric fit if the probability of falsely rejecting the circular fit is smaller than 5\% (P$_{\rm c} < 0.05$). The P$_{\rm c}$ values for all systems are given in Table\,\ref{tb:orbital_parameters}. Only one system, J053939.1--283329, has a circular orbit.

The resulting parameters and their errors are given in Table\,\ref{tb:orbital_parameters}. The best fitting Keplerian orbits are shown in Figs.\,\ref{fig:rv_PB6355} to \ref{fig:rv_PG2148+095}.

\begin{table*}
 \setlength\tabcolsep{1.5mm}
   \centering
   \caption{Orbital parameters derived from the radial velocity curves for the eleven new sdB binaries. The last column given the probability of falsly rejecting a circular orbit (P$_{\rm c}$) as defined by the Lucy \& Sweeney eccentricity test.} \label{tb:orbital_parameters}
   \begin{tabular}{lr@{ $\pm$ }lr@{ $\pm$ }lr@{ $\pm$ }lr@{ $\pm$ }lr@{ $\pm$ }lr@{ $\pm$ }lr@{ $\pm$ }lr@{ $\pm$ }lr}
    \hline   
    Object   &   \multicolumn{2}{c}{$P$}   &   \multicolumn{2}{c}{$T_0$}   &   \multicolumn{2}{c}{$e$}   &   \multicolumn{2}{c}{$\omega$}   &   \multicolumn{2}{c}{$K_{\rm{MS}}$}   &   \multicolumn{2}{c}{$\gamma_{\rm{MS}}$}   &   \multicolumn{2}{c}{$K_{\rm{sdB}}$}   &   \multicolumn{2}{c}{$\gamma_{\rm{sdB}}$}   &   \multicolumn{1}{c}{P$_{\rm c}$} \\
             &   \multicolumn{2}{c}{(d)}   &   \multicolumn{2}{c}{--2450000}   &   \multicolumn{2}{c}{}   &   \multicolumn{2}{c}{}   &   \multicolumn{2}{c}{(km/s)}   &   \multicolumn{2}{c}{(km/s)}  
             &   \multicolumn{2}{c}{(km/s)}   &   \multicolumn{2}{c}{(km/s)}  &  \multicolumn{1}{c}{( \% )} \\\hline \hline
    
    PB\,6355          &   684    &  31   &  7390   &  50   &  0.22   &  0.06   &  3.6   &  0.3             &  5.2    &  0.4   &  -1.3    &  0.2  &  16.1  &  0.5   &   1.7   &  0.2  &  0.02  \\
    MCT\,0146-2651    &   768    &  11   &  6890   &  104  &  0.08   &  0.06   &  5.6   &  0.9             &  6.6    &  0.5   &   39.7   &  0.2  &  10.0  &  0.8   &   41.4  &  0.3  &  1.40  \\
    FAUST\,321        &   993    &  15   &  5440   &  42   &  0.10   &  0.03   &  0.5   &  0.3             &  6.0    &  0.2   &  -38.6   &  0.1  &  13.2  &  0.2   &  -36.8  &  0.2  &  1.10  \\
    JL\,277           &   1082   &  9    &  5870   &  17   &  0.15   &  0.04   &  3.7   &  0.4             &  6.6    &  0.3   &  102.4   &  0.3  &  15.4  &  0.2   &  104.5  &  0.4  &  $<$0.01  \\
    J022836.7--362543 &   554    &  1    &  7131   &  8    &  0.15   &  0.02   &  5.0   &  0.1             &  9.2    &  1.4   &  -7.3    &  0.7  &  18.3  &  0.2   &   -2.1  &  0.2  &  $<$0.01  \\
    EC\,03143--5945   &   1037   &  3    &  7190   &  39   &  0.06   &  0.02   &  5.3   &  0.2             &  6.9    &  0.3   &  39.9    &  0.2  &  16.9  &  0.2   &   42.0  &  0.3  &  0.01  \\
    J033216.7--023302 &   1247   &  30   &  7332   &  53   &  0.18   &  0.05   &  3.8   &  0.2             &  6.6    &  1.1   &  23.7    &  0.7  &  18.1  &  1.7   &   29.6  &  1.1  &  2.30  \\
    J053939.1--283329 &   865    &  6    &  4876   &  21   &\multicolumn{2}{c}{0}& \multicolumn{2}{c}{/}   &  7.9    &  0.6   &  13.7    &  0.4  &  10.7  &  1.0   &   17.4  &  0.8  &  52  \\
    PG\,1514+034      &   479    &  2    &  6245   &  19   &  0.10   &  0.02   &  3.3   &  0.2             &  10.4   &  0.3   &  -72.0   &  0.2  &  17.8  &  0.4   &  -70.5  &  0.6  &  $<$0.01  \\
    J162842.0+111838  &   1176   &  30   &  6653   &  798  &  0.15   &  0.05   &  5.1   &  0.9             &  3.4    &  0.3   &  -43.0   &  0.3  &  8.1   &  0.5   &  -40.6  &  0.7  &  3.30  \\
    PG\,2148+095      &   1404   &  92   &  5979   &  108  &  0.21   &  0.06   &  4.9   &  0.8             &  7.1    &  0.9   &  -141.5  &  0.3  &  21.0  &  1.8   &  -138.6 &  0.3  &  0.02  \\\hline
    
   \end{tabular}
\end{table*}

\section{The Periods and mass ratios of wide sdB binaries}

Currently there are eleven long period sdB+MS binaries with solved orbits. Of those systems ten have a known mass ratio; PG\,1018--047 \citep{Deca2012, Deca2018}, PG\,1104+243 \citep{Vos2012, Barlow2012}, PG\,1449+653 \citep{Barlow2013}, Feige\,87 \citep{Barlow2012, Vos2013}, BD+34$^{\circ}$1543, BD+29$^{\circ}$3070 \citep{Vos2013} and lastly BD--7$^{\circ}$5977, EC\,11031--1348, TYC\,2084--448--1 and TYC\,3871--835--1 \citep{Vos2017}. Furthermore the orbital period of PG\,1701+359 was determined by \citet{Barlow2013}, but radial velocities for the companion could not be derived from the spectra, thus the mass ratio is not known. Lastly there is one system, EC\,20117--4014 of which an orbital period is determined based on the light curve analysis by \citet{Otani2018}. Together with the eleven systems presented in this article, there are now 23 wide sdB binaries for which the orbital period  is known, and 21 of them have fully solved orbits for both components. The orbital periods and mass ratios of all these systems are summarized in Table\,\ref{tb:p_q_summary}. For those systems for which the mass ratio is known, we have calculated the orbital separation using the sdB mass derived from the orbital period - sdB mass relation of \citet{Chen2013} (see Sect.\,\ref{s:rlof_stability} and Table.\,\ref{tb:sdb_masses}).

\begin{table}
 \centering
   \caption{The orbital period, separation and mass ratio of all 23 known long period sdB binaries with solved orbits. The separation is calculated using the sdB mass derived from the orbital period - sdB mass relation of \citet{Chen2013}, see Table\,\ref{tb:sdb_masses}.} \label{tb:p_q_summary}
   \begin{tabular}{lr@{ $\pm$ }lr@{ $\pm$ }lr@{ $\pm$ }l}
   \hline\hline
   Object   &   \multicolumn{2}{c}{$P$} &    \multicolumn{2}{c}{$a$} &   \multicolumn{2}{c}{$q$} \\
            &   \multicolumn{2}{c}{(d)} &    \multicolumn{2}{c}{(AU)} &   \multicolumn{2}{c}{$M_{\rm sdB} / M_{\rm MS}$} \\\hline
   PG\,1514+034            &  479 &  2 & 1.23 & 0.01 & 0.58 & 0.03 \\
   J022836.7--362543       &  554 & 10 & 1.42 & 0.06 & 0.50 & 0.08 \\
   PB\,6355                &  684 & 31 & 1.84 & 0.07 & 0.32 & 0.02 \\
   PG\,1701+359            &  734 & 15 & \multicolumn{2}{c}{/} & \multicolumn{2}{c}{/} \\
   PG\,1018--047           &  752 &  2 & 1.65 & 0.01 & 0.70 & 0.02 \\
   PG\,1104+243            &  755 &  3 & 1.65 & 0.01 & 0.71 & 0.02 \\
   MCT\,0146-2651          &  768 & 11 & 1.69 & 0.02 & 0.66 & 0.03 \\
   EC\,20117-4014          &  795 &  1 & \multicolumn{2}{c}{/} & \multicolumn{2}{c}{/} \\
   J053939.1--283329       &  865 & 20 & 1.81 & 0.04 & 0.74 & 0.09 \\
   PG\,1449+653            &  909 &  2 & 1.88 & 0.05 & 0.73 & 0.10 \\
   Feige\,87               &  938 &  2 & 2.04 & 0.01 & 0.55 & 0.01 \\
   BD+34$^{\circ}$1543     &  972 &  2 & 2.08 & 0.01 & 0.57 & 0.01 \\
   FAUST\,321              &  993 & 15 & 2.22 & 0.03 & 0.45 & 0.01 \\
   EC\,03143--5945         & 1037 & 10 & 2.35 & 0.03 & 0.41 & 0.02 \\
   JL\,277                 & 1082 &  9 & 2.42 & 0.04 & 0.42 & 0.02 \\
   TYC\,2084--448--1       & 1098 &  5 & 2.33 & 0.02 & 0.51 & 0.02 \\
   EC\,11031--1348         & 1099 & 12 & 2.53 & 0.04 & 0.36 & 0.02 \\
   J162842.0+111838        & 1167 & 13 & 2.54 & 0.05 & 0.42 & 0.04 \\
   J033216.7--023302       & 1247 & 30 & 2.77 & 0.15 & 0.36 & 0.07 \\
   BD+29$^{\circ}$3070     & 1254 &  5 & 2.77 & 0.04 & 0.37 & 0.02 \\
   BD--7$^{\circ}$5977     & 1262 &  2 & 2.67 & 0.16 & 0.44 & 0.10 \\
   TYC\,3871--835--1       & 1263 &  3 & 2.55 & 0.02 & 0.54 & 0.02 \\
   PG\,2148+095            & 1404 & 92 & 3.08 & 0.20 & 0.34 & 0.06 \\
   \hline
   \end{tabular}
\end{table}

\subsection{The effect of the RV detection limits on the observed sample.}\label{s:observational_bias}
Before we consider the properties of the observed sample, it is important to understand the effects of the radial velocity detection limits of our observing program on these properties. Based on the accuracy of the different spectrographs (UVES, FEROS, CHIRON and HERMES), and the methods used to determine the radial velocities of both the sdB and the MS components we can detect and solve orbits with radial velocity amplitudes at least down to K$_{\rm sdB}$ $\geq$ 5 \kms\ and K$_{\rm MS}$ $\geq$ 3 \kms. In fact, the lowest detected amplitudes are for the system TYC\,3871--835--1 observed with the HERMES spectrograph, with K$_{\rm sdB}$ = 4.24 $\pm$ 0.20 \kms\ and K$_{\rm MS}$ = 2.31 $\pm$ 0.04 \kms. Up to now we have not observed any sdB+FGK system in which we could not detect any radial velocity variations.

To investigate the effect of these detection limits on the observed periods and mass ratios, we calculated which percentage of the possibly existing systems we can observe for different period - mass ratio combinations. As we are only interested in finding the effect of the detection limits, no stellar or binary evolution is taken into account in this procedure. To calculate the fraction of systems that we observed compared to the total number of systems that can exist, the sdB and MS radial velocity amplitudes for a random sample of systems are calculated. The properties of these systems were selected as follows. The mass of the sdB star is picked between 0.35 and 0.55 M$_{\odot}$, while the mass of the MS star varies between 0.60 and 1.50 M$_{\odot}$. The orbital period and eccentricity are limited between respectively 400 and 1400 days and 0.0 - 0.30 to cover the ranges of the observed sample. Finally, the inclination is randomly selected on the sphere, with the fraction of randomly oriented axes between $i$ and $i + {\rm d}i$ being proportional to:
\begin{equation}
 P(i){\rm d}i = \sin{i}\ {\rm d}i
\end{equation}
The radial velocity amplitudes are calculated using:
\begin{equation}
  K_{1,2} = \sqrt[3\,]{\frac{M_{2,1}^3\ \sin{i}^3}{(M_1 + M_2)^2}\ \frac{2\,\pi\,G}{P}\ \frac{1}{(1-e^2)^{3/2}}}
\end{equation}

\begin{figure}
    \includegraphics{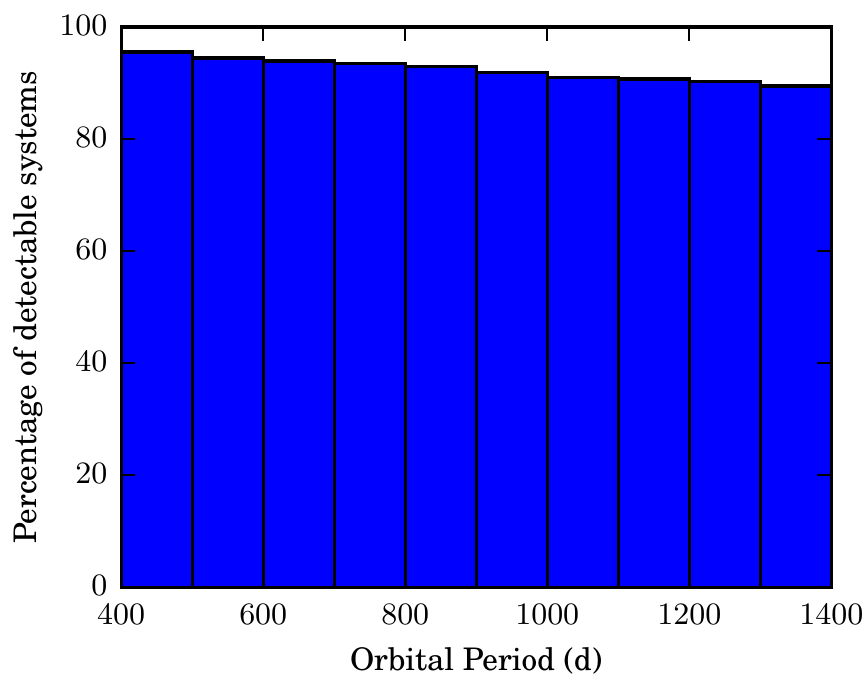}
    \caption{The percentage of all sdB+MS systems in different period bins that can be detected with our observing setup with limits K$_{\rm sdB}$ $\geq$ 5 \kms\ and K$_{\rm MS}$ $\geq$ 3 \kms.}
    \label{fig:f_detection_limits_P}
\end{figure}

First, we can check which fraction of systems that are observable as a function of orbital period. Systems with longer orbital periods have lower amplitudes, and are thus harder to observe. In Fig.\,\ref{fig:f_detection_limits_P} we show the fraction of sdB+MS systems observable as a function of the orbital period. We find that at the short period end at 400 days we can detect 96\,\% of the systems, while at longer periods around 1400 days we can detect $\sim$ 90\,\% of all possible systems.

\begin{figure}
    \includegraphics{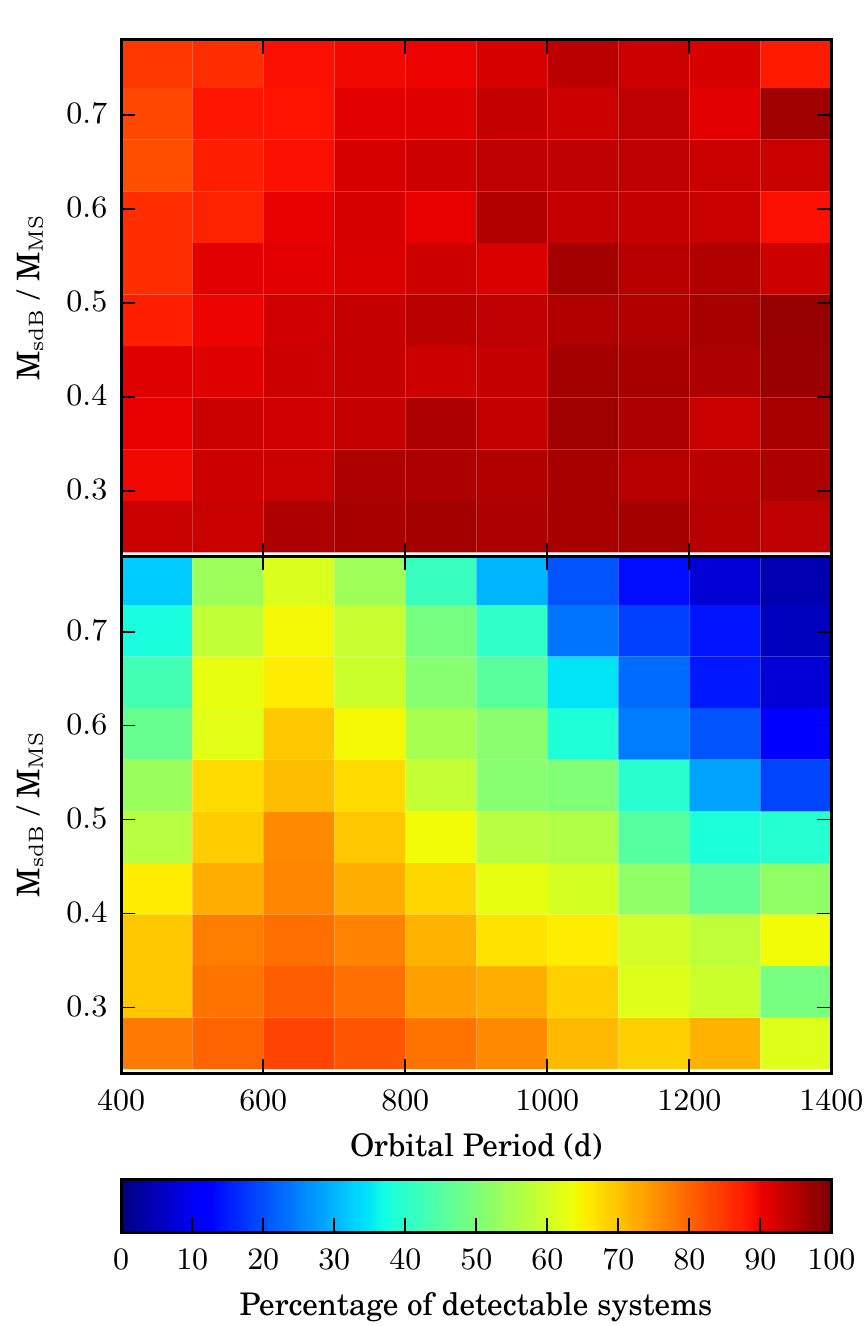}
    \caption{The percentage of all sdB+MS systems in each period - mass ratio bin that can be detected with our observing setup depending on the orbital period and mass ratio. Top panel: real detection limit of K$_{\rm sdB}$ $\geq$ 5 \kms\ and K$_{\rm MS}$ $\geq$ 3 \kms. Bottom panel: higher detection limits of K$_{\rm sdB}$ $\geq$ 14 \kms\ and K$_{\rm MS}$ $\geq$ 5 \kms. Higher detections limits could cause a period - mass ratio relation in which systems at longer periods and higher mass ratios would not be detectable. However, our actual detection limit does not lead to any pattern in period - mass ratio.}
    \label{fig:f_detection_limits_P_q}
\end{figure}

Due to its construction this sample has no relation between mass ratio and orbital period. Using the detection limits, the fraction of systems that can be observed can be calculated for different period - mass ratio combinations. This is shown in the top panel of Fig.\,\ref{fig:f_detection_limits_P_q}. As can be seen in that figure, our detection limits do not cause a significant correlation between orbital period and mass ratio. The minimum percentage of detectable systems is 83\,\% for the highest mass ratios irrespective of the orbital period, while the maximum detection rate is around 98\,\% at low mass ratios and average orbital periods. If our detection limits would be higher, e.g. K$_{\rm sdB}$ $\geq$ 14 \kms and K$_{\rm MS}$ $\geq$ 5 \kms, these limits would prevent us from observing a significant part of the period - mass ratio (P-q) distribution, and would result in a correlation between mass ratio and orbital period. This is shown in the bottom panel of Fig.\,\ref{fig:f_detection_limits_P_q}.

Based on this analysis we can conclude that the observed distribution of orbital periods is only slightly affected by the detection limits of our observing program. Therefore, any observed relation between the orbital period and the mass ratio has to be caused by evolutionary effects. 

\subsection{The effect of target selection on the observed sample.}\label{s:selection_bias}
The easiest systems to identify from both spectral and color selection are sdB binaries with an early F type companion as in that case both the sdB and the F star are clearly visible in the spectrum, and the F star would cause a clear IR-excess which is visible in the 2MASS colors. In our sample the binaries have cool companions ranging from mid-K to mid-F type MS stars. Systems in which the cool companion would be M or K5-9 type stars could potentially not have been selected as the cool star might not have been visible. However, K5 and earlier type stars have masses lower than the expected mass of an sdB star ($\sim 5$ M$_{\odot}$). In this case, there is no possibility for the mass-loss phase to be stable, and a common envelope with a spiral in phase will take place. Therefor these systems are not expected to be part of the wide sdB+MS population.

On the other end, it is possible that systems with a late type companion (F5-A) could have been missed. As the likely progenitors of the sdB binaries had masses up to $\sim$1.8 M$_{\odot}$, it is possible that sdB stars with companions up to early A-type stars could form. In this case, it is possible that they are not recognized as containing an sdB component. In practice, the composite sdB binary sample presented in \citet{Vos2018a} on which this survey is based does not contain sdB+A type binaries. It is thus possible that sdB binaries with companions varying from mid-F and early-A type stars exist, but are not included in our sample. The possible effects of this are discussed in the following sections.

\subsection{The orbital period distribution}\label{s:period-distribution}
Building on the work of \citet{Han2002, Han2003}, \citet{Chen2013} performed a binary population synthesis (BPS) study focused on sdB binaries produced by the first stable RLOF channel. The orbital periods of sdB+MS systems depended on the mass of the sdB progenitors, with a division between progenitors with a mass higher than 2 M$_{\odot}$ and those less than 2 M$_{\odot}$. Systems that had a progenitor mass $<$ 2 M$_{\odot}$ show a strong dependence between final sdB mass and orbital period, with orbital periods ranging from 400 to 1100 days. For systems with a progenitor mass higher than 2 M$_{\odot}$ the final orbital period depends strongly on the angular momentum loss, and these systems are predicted to have orbital periods ranging from a few days up to $\sim$ 100 days. The division between these two groups is caused by the sharp drop of the radius of RGB stars with ZAMS masses below 2 M$_{\odot}$ and H-rich envelope mass lower than a certain value relative to stars with ZAMS masses over 2 M $_{\odot}$.

\begin{figure}
    \includegraphics{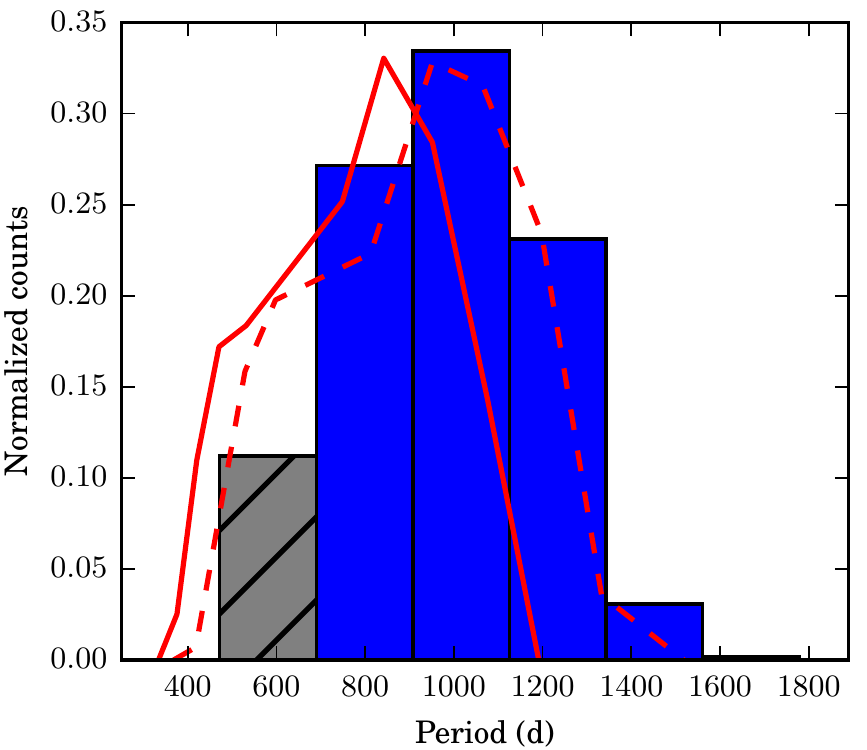}
    \caption{The distribution of orbital periods of all solved wide sdB binaries as described in Sect.\,\ref{s:period-distribution}. Blue full bars: the distribution of the systems following the main p-q trend, gray hatched bar: including the three systems laying underneath the p-q relation. The predicted period distribution of \citet{Chen2013} is shown in red. Full line: no atmospheric RLOF, dashed line: including atmospheric RLOF. The predicted period distribution is rescaled to take into account the difference in bin size between the observed and theoretical distribution.}
    \label{fig:period_distribution}
\end{figure}

The period distribution of the observed wide sdB+MS sample is shown in Fig.\,\ref{fig:period_distribution}. This distribution takes into account the errors on the observed periods by treating each observation as a normal distribution with the error as standard deviation. In practice this distribution is obtained by picking 1000 orbital period from a Gaussian distribution centered at the observed period with the error on the observed period as standard deviation. This is done for all systems, and the histogram shown in Fig.\,\ref{fig:period_distribution} is created from these 23000 points. To avoid over interpreting the data, the bins of the histogram are set at 200 days. The distribution of all systems that are part of the period -- mass ratio relation is shown in blue, while the contribution of the three outlier systems is shown in gray (see also Sect.\,\ref{s:pq_relation}).

The periods of all observed systems are significantly longer than those predicted for sdB+MS binaries originating from high-mass sdB progenitors. However, they match well with the predictions for sdB+MS binaries with low-mass progenitors (red full line in Fig.\,\ref{fig:period_distribution}). The shortest orbital period of 479 days matches with the lower limit of 400 days from the BPS models, but the longer period systems do reach periods a few 100 days longer than predicted. A solution proposed by \citet{Chen2013} is that their calculations do not take atmospheric RLOF into account, which would increase the final orbital period. When atmospheric RLOF is included (red dashed line in Fig.\,\ref{fig:period_distribution}), orbital periods as long as $\sim$ 1600 days can be reached, thus covering all observed systems.

If the three outlying systems would be the result of another formation channel, they should be excluded from this comparison (thus only the blue bars in Fig.\,\ref{fig:period_distribution} should be considered). In that case the main conclusion still holds. The observed period distribution matches with the predictions of the BPS studies of \citet{Chen2013}. Only the lower limit of the predicted period range is a few 100 days shorter than the lower end of the observed periods. Following the results of \citet{Chen2013} that the final orbital period is directly dependent on the final mass of the sdB star, this would limit the mass of the sdB stars between 0.40 (Z = 0.02) and 0.50 M$_{\odot}$ (Z = 0.004) If the three outlying systems are excluded, the lower mass limit is 0.43 M$_{\odot}$ (Z = 0.02), see also Sect.\,\ref{s:rlof_stability}.

\subsection{The period - mass ratio relation}\label{s:pq_relation}
\begin{figure*}
    \includegraphics{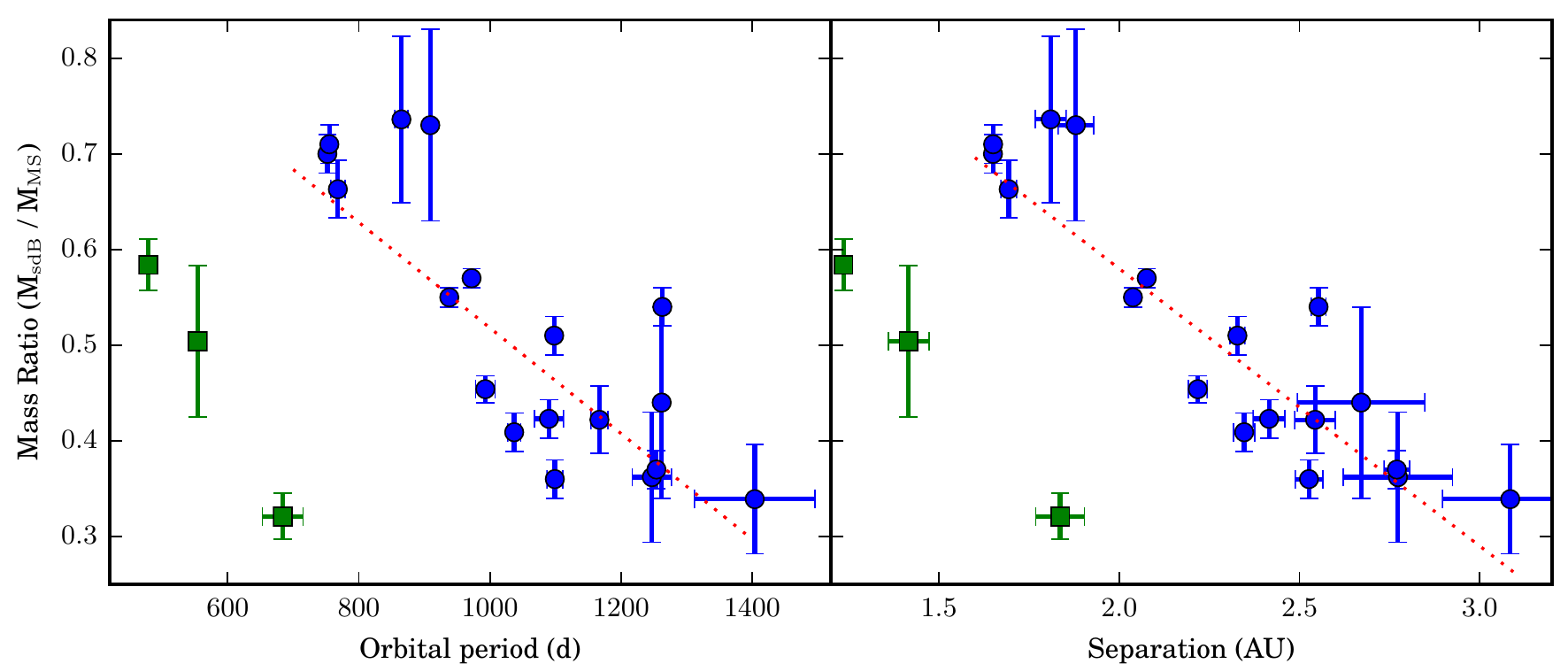}
    \caption{The mass ratio as a function of orbital period (left) and separation (right) of 21 wide sdB binaries with known orbital parameters. The separation is calculated using the sdB mass derived from the orbital period - sdB mass relation of \citet{Chen2013}, see Table\,\ref{tb:sdb_masses}. There are two separate groups visible. The main group (blue circles) follows a strong relation of a lower mass ratio at longer orbital periods. These systems are thought to be formed from progenitors with a degenerate He core. For this group, the best fitting linear relations between the mass ratio and the orbital period and separation given in Eq.\,\ref{eq:p_q} and \ref{eq:a_q} are plotted in red dotted line. The second group consisting of 3 systems (green squares) has shorter orbital periods than the main group, and are thought to be formed from more massive progenitors with a non degenerate He core.}
    \label{fig:p_a_q_relation}
\end{figure*}

When comparing the orbital periods and the mass ratios of the wide sdB binaries it is immediately clear that there is a strong correlation between those two observables. In Fig.\,\ref{fig:p_a_q_relation} the mass ratio is plotted versus the orbital period on the left and versus the orbital separation on the right. The mass ratio is defined as $q = M_{\rm sdB} / M_{\rm MS}$.  It is clear from those figures that the majority of the systems (18 out of 21, shown in blue circles in Fig.\,\ref{fig:p_a_q_relation}) show a clear relation of higher mass ratios at shorter orbital periods. The Pearson test \citep{Pearson1896} was used to calculate the correlation between these observables. It is defined as the covariance of the two variables divided by their standard deviation. In the case of the period versus the mass ratio, the Pearson test yields a negative correlation of $r(P,q) = -0.83$ with a confidence level larger than 99\% ($p$ value $<$ 0.001). In the case of the orbital separation versus the mass ratio, the correlation is even stronger with of $r(a,q) = -0.91$ with $p$ value $<$ 0.001. A linear fit to the P-q and separation - q relation yields the following observed relations, with errors in brackets:
\begin{align}
 P(q) &= -5.52\,(1.12)\cdot10^{-4} \cdot q + 1.07\,(0.11)\ {\rm d} \label{eq:p_q}\\ 
 a(q) &= -0.29\,(0.04) \cdot q + 1.16\,(0.08)\ {\rm AU} \label{eq:a_q}
\end{align}
Only the systems in the main group of the P-q and a-q diagram (blue circles in Fig.\,\ref{fig:p_a_q_relation}) are taken into account when calculating these linear relations. They are plotted on Fig\,\ref{fig:p_a_q_relation} in red dotted lines.

There are three systems -- PG\,1514+034, J022836.7--362543 and PB\,6355 (shown in green squares in Fig.\,\ref{fig:p_a_q_relation}) -- that do not follow this $P-q$ relation. It can be argued that these three outliers show the same trend of higher mass ratios at shorter orbital periods, but shifted to lower orbital periods. However, as there are only three systems in this secondary group, a quantitative statement about a possible relation cannot be made. It is however clear that there are two distinct groups with a gap in between them. At this point, taking into account the observational bias (see Sect.\,\ref{s:observational_bias}), it would be unjustified to assume that there would be a continuous distribution and that the gap between the two groups will be filled when more systems will be observed. 

As mentioned in Sect.\,\ref{s:selection_bias}, the wide sdB sample might be lacking sdB binaries with late F to A-type companions. Based on the correlation between the orbital period and the mass ratio, these systems would be located at the long period end of the P-q relation. It is thus possible that future observations will extend the P-q relation to longer orbital periods. 

\section{The stability of RLOF}\label{s:rlof_stability}
According to the sdB mass - orbital period relation given by \citet{Chen2013}, we obtained the masses of sdBs with well-known orbital periods, as shown in Table \ref{tb:sdb_masses}. For simplicity, we only consider population I stars, and the atmospheric-RLOF is included to cover the systems with orbital periods longer than $\sim 1200$ days. The minimum and maximum sdB masses are obtained from the observational errors on the orbital periods. 

In Table \ref{tb:sdb_masses}, we see that the sdB mass for the first eight systems is less than $0.4425\,M_{\odot}$, the minimum sdB mass obtained from the stable RLOF channel when the progenitor of the sdB star has a mass of $M_1=1.6\,M_{\odot}$ \citep[Table 4 in][]{Han2002}. It means that these objects are likely produced from progenitors with $M_1>1.6M_{\odot}$, where the helium core in giants transfers from degenerate state to non-degenerate state  gradually. In this mass range, with an increase of the stellar mass, the core mass range for He ignition decreases first and increases again after $M_1$ is more massive than $\sim 2M_\odot$, as shown in Fig.\,1 of 
\citet{Han2002}\footnote{The core mass ranges for He ignition shown in Fig.\,1 of \citet{Han2002} are for products from common envlope ejection. The minimum He ignition core mass after stable RLOF is a little different from that of common envelope ejection due to the different core mass increasing processes in giant stars, but the difference is little (see the sdB mass from Table\,\ref{tb:sdb_masses} and the minimum core mass for He ignition in Table 1 in \citet{Han2002}).} 
So, there is a minimum in the parameter space for producing sdB stars when the He core changes from degenerate to non-degenerate state. The gap near the short-orbital period end in Fig.\,\ref{fig:p_a_q_relation} then could be understood and the outliers in the Figure (i.e. the first three samples in Table\,\ref{tb:p_q_summary}) are likely from more massive progenitors with non-degenerate He cores. Detailed binary evolution calculations are necessary to confirm this, and the mass of the outliers should be revisited accordingly. 
 
\begin{figure}
    \includegraphics[width=\columnwidth]{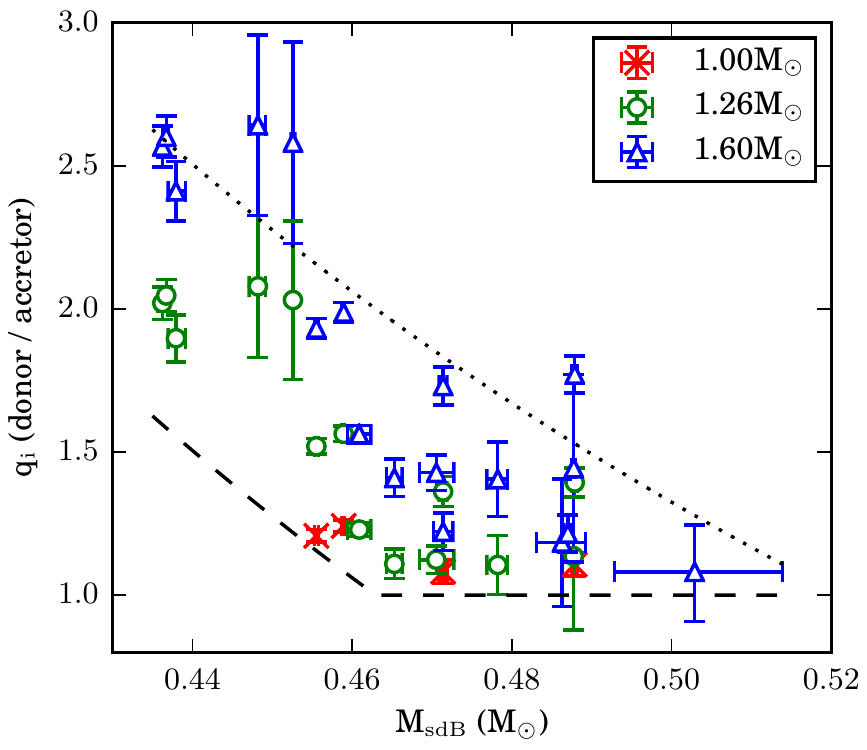}
    \caption{The initial mass ratio at the start of RLOF in function of the mass of the sdB star (which is roughly the core mass of its progenitor at the start of RLOF). Only the systems with an sdB progenitor mass lower than 2.0 M$_{\odot}$ are shown. The initial mass ratio is calculated assuming no material has been accreted by the companion during RLOF. The different symbols indicate different possible progenitor masses of 1.00 M$_{\odot}$ (red crosses), 1.26 M$_{\odot}$ (green circles) and 1.60 M$_{\odot}$ (blue triangles). Only cases with q$_{\rm i} > 1.0$ are shown. 
    The dotted and dashed lines indicate respectively the upper and lower limits for the initial mass ratio and are given by Equations \ref{eq:qinit_high} and \ref{eq:qinit_low} in the text.}
    \label{fig:initial_q}
\end{figure}

Given the sdB mass in Table \ref{tb:sdb_masses} and the mass ratio of Table\,\ref{tb:p_q_summary}, for each system, 
we can calculate the companion mass and corresponding errors, from which the initial companion mass can be derived for various accretion efficiencies during RLOF. Eventually, we can obtain the mass ratio at the onset of RLOF for a given progenitor mass of sdB stars. The results are shown in Fig.\,\ref{fig:initial_q}, where 5 systems (the first 3 in Table 3, and the two without observed mass ratio PG\,1701+359 and EC\,20117--4014) have been excluded.

We assume that the companion has not accreted any material during RLOF, which gives the maximum initial companion mass and the lowest mass ratio of the binary at the onset of RLOF. This assumption is supported by estimates of the amount of accreted mass based on the rotational velocity of the companions \citep{Vos2018a}, which indicates that very little mass is accreted by the companions during the RLOF phase. Further support for this assumption is that the evolutionary state of some of the companions does not allow for the accretion of large amounts of mass \citep[e.g.][]{Vos2012}. Although the assumption of no accretion might seem somewhat ad-hoc, the derived $q_{\rm i}$ ($M_{\rm prog} / M_{\rm comp}$) in Fig.\,\ref{fig:initial_q} is the lower limit for the mass ratio at the onset of RLOF, since the companion is less massive if accretion is included. Furthermore, only the models with initial mass ratios greater than 1.0 have been presented in the figure, since the progenitors of sdB stars are more evolved and should be initially more massive than their MS companions. 

As shown in Fig.\,\ref{fig:initial_q}, the initial mass ratio generally decreases with sdB mass. As the hydrogen envelope is only about 1-2\,\% of the total mass of an sdB star, the sdB mass corresponds roughly to the core mass of the red giant progenitor at the start of RLOF. Different progenitor masses give different initial mass ratios as expected. We can define an upper limit on the initial mass ratio (q$_{\rm i}$) in function of the sdB mass above which RLOF will not be stable anymore:
\begin{equation}
 q_{\rm i, high} = M_{\rm sdB}^{-2}-0.25\,M_{\rm sdB}-2.55, \label{eq:qinit_high}
\end{equation}
where M$_{\rm sdB}$ is the mass of sdB star in solar mass. This relation is shown in dotted black line in Fig.\,\ref{fig:initial_q}. The upper boundary of the mass ratios seen in Fig.\,\ref{fig:p_a_q_relation} can be well reproduced, if the dotted line in Fig.\,\ref{fig:initial_q} is used as the critical mass ratio for stable RLOF. Moreover, the critical mass ratio given by the dotted line is consistent with that obtained from detailed binary evolution calculations performed by \citet{Han2002} and \citet{Chen2008}, that is, the critical mass ratio decreases when the giant evolves upwards along the giant branch and its core mass increases.

In a similar fashion a lower limit on the initial mass ratio can be derived as:
\begin{equation}
 q_{\rm i, low}  = {\rm Max}(\ M_{\rm sdB}^{-2}-0.25\,M_{\rm sdB}-3.55\ ,\ 1.0\ ), \label{eq:qinit_low}
\end{equation}
where the minimum mass ratio has to be larger than one. This relation is shown in dashed black line in Fig.\,\ref{fig:initial_q}. Applying Equation\,\ref{eq:qinit_low} as a lower limit on the inital mass ratio will reproduce the lower boundary of the mass ratios in Fig.\,\ref{fig:p_a_q_relation}. However, at this moment such a lower limit is not predicted by theoretical models.

sdB binaries with late F to A type companions are expected at longer orbital period. The current sample does not contain any of these system, but by extrapolating the observed relations their location in Fig.\,\ref{fig:initial_q} can be estimated. sdB+A binaries would have initial mass ratios close to 1, and following the relationd from \citet{Chen2013} have higher core masses. Henceforth they would fill in the lower right corner of Fig.\,\ref{fig:initial_q}.

\begin{table}
 \centering
   \caption{The sdB masses for all 23 wide sdB binaries with known orbital periods obtained from the sdB mass - period relation given by \citet{Chen2013} for population I. The minimum and maximum sdB masses are obtained from observational errors on the orbital periods. Atmoshperic-RLOF is included.} \label{tb:sdb_masses}
   \begin{tabular}{lccc}
   \hline\hline
   Object   &   \multicolumn{1}{c}{$M_{\rm sdB}$} &    \multicolumn{1}{c}{$M_{\rm sdB}^{\rm min}$} &   \multicolumn{1}{c}{$M_{\rm sdB}^{\rm max}$} \\
            &   \multicolumn{1}{c}{(M$_{\odot}$)} &    \multicolumn{1}{c}{(M$_{\odot}$)} &   \multicolumn{1}{c}{(M$_{\odot}$)} \\\hline
 PG\,1514+034        & 0.4038 & 0.4035 & 0.4040 \\
 J022836.7--362543   & 0.4135 & 0.4123 & 0.4147 \\
 PB\,6355            & 0.4289 & 0.4255 & 0.4323 \\
 PG\,1701+359        & 0.4344 & 0.4328 & 0.4360 \\
 PG\,1018--047       & 0.4362 & 0.4361 & 0.4365 \\
 PG\,1104+243        & 0.4367 & 0.4362 & 0.4370 \\
 MCT\,0146-2651      & 0.4379 & 0.4370 & 0.4392 \\
 EC\,20117-4014      & 0.4409 & 0.4409 & 0.4410 \\
 J053939.1--283329   & 0.4482 & 0.4471 & 0.4492 \\
 PG\,1449+653        & 0.4526 & 0.4523 & 0.4528 \\
 Feige\,87           & 0.4555 & 0.4553 & 0.4558 \\
 BD+34$^{\circ}$1543 & 0.4589 & 0.4587 & 0.4591 \\
 FAUST\,321          & 0.4609 & 0.4594 & 0.4624 \\
 EC\,03143--5945     & 0.4653 & 0.4643 & 0.4663 \\
 JL\,277             & 0.4705 & 0.4685 & 0.4727 \\
 TYC\,2084--448–1    & 0.4714 & 0.4709 & 0.4719 \\
 EC\,11031--1348     & 0.4714 & 0.4702 & 0.4726 \\
 J162842.0+111838    & 0.4782 & 0.4769 & 0.4794 \\
 J033216.7--023302   & 0.4863 & 0.4831 & 0.4892 \\
 BD+29$^{\circ}$3070 & 0.4870 & 0.4865 & 0.4875 \\
 BD--7$^{\circ}$5977 & 0.4877 & 0.4875 & 0.4880 \\
 TYC\,3871--835--1   & 0.4879 & 0.4875 & 0.4882 \\
 PG\,2148+095        & 0.5029 & 0.4929 & 0.5139 \\
   \hline
   \end{tabular}
\end{table}

\section{Summary and Conclusions}
In this article we have presented orbital solutions for 11 new long period hot subdwarf binaries with main sequence companions based on spectroscopic observations taken with the UVES, FEROS and CHIRON spectrographs. These systems are all part of a long term observing program focused on wide composite sdB binaries. This brings the total number of systems with solved orbital parameters to 21, with two more systems for which orbital periods are known, but for which no mass ratio could be derived.

An analysis of the detection limits of our observing program shows that the sensitivity of the observations is sufficiently high that it does not impose any correlations on the observed orbital properties. All observed relations are thus caused by evolutionary effects.

The orbital period distribution corresponds very well to the predicted period distribution for composite sdB binaries formed through the first stable RLOF formation channel of \citet{Chen2013}. The observed orbital period distribution clearly indicates that including atmospheric RLOF is necessary to explain the very long period systems. 

An important new result is the discovery of a strong correlation between orbital period and the mass ratio. In the P-q plane, two groups of systems are visible, with the majority (18 systems) following a tight relation of lower mass ratio at longer orbital periods. The remaining three systems form a separate group that is located at shorter orbital periods with respect to the main group. The second group shows a similar P-q relation as the main group, but with only three systems this is not statistically significant. 

The observed P-q relation can be linked to the stability of RLOF on the RGB. Assuming that the sdB mass follows the sdB mass - period relation obtained by \citet{Chen2013} and that no mass has been accreted by the companion during RLOF, we show the initial mass ratio of the observed long period sdB samples depends on the progenitor core mass at the onset of RLOF. More specifically, the initial mass ratio decreases with increasing core mass, which is in accordance with the theoretical results of \citet{Chen2008} that show that the critical mass ratio for stable RLOF decreases when the giant donor evolves upwards along the giant branch and its core mass increases. Based on our observations we derived an upper and lower limit on the initial mass ratio in function of the core mass, in between which a binary system will undergo stable RLOF.

\section*{Acknowledgements}
JV acknowledges financial support from FONDECYT in the form of grant number 3160504.
XC and ZH acknowledge financial support from the Natural Science Foundation of China (Nos. 11733008, 11521303) and from the Yunnan province (No.2017HC018).
TB and BB acknowledge High Point University for providing the funds necessary to purchase observing time on the CTIO/SMARTS 1.5-m telescope through the SMARTS Consortium.
Based on observations collected at the European Organisation for Astronomical Research in the Southern Hemisphere under ESO programmes 088.D-0364(A), 093.D-0629(A), 096.D-0180(A), 097.D-0110(A), 098.D-0018(A), 099.D-0014(A), 099.A-9019(A) and 0100.D-0082(A). 
Based on observations obtained at the CTIO/SMARTS 1.5-m telescope through the CNTAC un program CN2018A-23.
This research has used the services of {\sc Astroserver.org} under reference KW32YZ and EZZ74T.

%%%%%%%%%%%%%%%%%%%%%%%%%%%%%%%%%%%%%%%%%%%%%%%%%%



\bibliographystyle{mnras}
\bibliography{bibliogaphy} % if your bibtex file is called example.bib

\begin{thebibliography}{}
\makeatletter
\relax
\def\mn@urlcharsother{\let\do\@makeother \do\$\do\&\do\#\do\^\do\_\do\%\do\~}
\def\mn@doi{\begingroup\mn@urlcharsother \@ifnextchar [ {\mn@doi@}
  {\mn@doi@[]}}
\def\mn@doi@[#1]#2{\def\@tempa{#1}\ifx\@tempa\@empty \href
  {http://dx.doi.org/#2} {doi:#2}\else \href {http://dx.doi.org/#2} {#1}\fi
  \endgroup}
\def\mn@eprint#1#2{\mn@eprint@#1:#2::\@nil}
\def\mn@eprint@arXiv#1{\href {http://arxiv.org/abs/#1} {{\tt arXiv:#1}}}
\def\mn@eprint@dblp#1{\href {http://dblp.uni-trier.de/rec/bibtex/#1.xml}
  {dblp:#1}}
\def\mn@eprint@#1:#2:#3:#4\@nil{\def\@tempa {#1}\def\@tempb {#2}\def\@tempc
  {#3}\ifx \@tempc \@empty \let \@tempc \@tempb \let \@tempb \@tempa \fi \ifx
  \@tempb \@empty \def\@tempb {arXiv}\fi \@ifundefined
  {mn@eprint@\@tempb}{\@tempb:\@tempc}{\expandafter \expandafter \csname
  mn@eprint@\@tempb\endcsname \expandafter{\@tempc}}}

\bibitem[\protect\citeauthoryear{{Baranne} et~al.,}{{Baranne}
  et~al.}{1996}]{Baranne1996}
{Baranne} A.,  et~al., 1996, \aaps, \href
  {http://adsabs.harvard.edu/abs/1996A%26AS..119..373B} {119, 373}

\bibitem[\protect\citeauthoryear{{Barlow}, {Wade}, {Liss}, {{\O}stensen}  \&
  {Van Winckel}}{{Barlow} et~al.}{2012}]{Barlow2012}
{Barlow} B.~N.,  {Wade} R.~A.,  {Liss} S.~E.,  {{\O}stensen} R.~H.,   {Van
  Winckel} H.,  2012, \mn@doi [\apj] {10.1088/0004-637X/758/1/58}, \href
  {http://adsabs.harvard.edu/abs/2012ApJ...758...58B} {758, 58}

\bibitem[\protect\citeauthoryear{{Barlow}, {Liss}, {Wade}  \& {Green}}{{Barlow}
  et~al.}{2013}]{Barlow2013}
{Barlow} B.~N.,  {Liss} S.~E.,  {Wade} R.~A.,   {Green} E.~M.,  2013, \mn@doi
  [\apj] {10.1088/0004-637X/771/1/23}, \href
  {http://adsabs.harvard.edu/abs/2013ApJ...771...23B} {771, 23}

\bibitem[\protect\citeauthoryear{{Beck} et~al.,}{{Beck}
  et~al.}{2014}]{Beck2014}
{Beck} P.~G.,  et~al., 2014, \mn@doi [\aap] {10.1051/0004-6361/201322477},
  \href {http://adsabs.harvard.edu/abs/2014A%26A...564A..36B} {564, A36}

\bibitem[\protect\citeauthoryear{{Brahm}, {Jord{\'a}n}  \& {Espinoza}}{{Brahm}
  et~al.}{2017}]{Brahm2017}
{Brahm} R.,  {Jord{\'a}n} A.,   {Espinoza} N.,  2017, \mn@doi [\pasp]
  {10.1088/1538-3873/aa5455}, \href
  {http://adsabs.harvard.edu/abs/2017PASP..129c4002B} {129, 034002}

\bibitem[\protect\citeauthoryear{{Chen} \& {Han}}{{Chen} \&
  {Han}}{2008}]{Chen2008}
{Chen} X.,  {Han} Z.,  2008, \mn@doi [\mnras]
  {10.1111/j.1365-2966.2008.13334.x}, \href
  {http://adsabs.harvard.edu/abs/2008MNRAS.387.1416C} {387, 1416}

\bibitem[\protect\citeauthoryear{{Chen}, {Han}, {Deca}  \&
  {Podsiadlowski}}{{Chen} et~al.}{2013}]{Chen2013}
{Chen} X.,  {Han} Z.,  {Deca} J.,   {Podsiadlowski} P.,  2013, \mn@doi [\mnras]
  {10.1093/mnras/stt992}, \href
  {http://adsabs.harvard.edu/abs/2013MNRAS.434..186C} {434, 186}

\bibitem[\protect\citeauthoryear{{Copperwheat}, {Morales-Rueda}, {Marsh},
  {Maxted}  \& {Heber}}{{Copperwheat} et~al.}{2011}]{Copperwheat2011}
{Copperwheat} C.~M.,  {Morales-Rueda} L.,  {Marsh} T.~R.,  {Maxted} P.~F.~L.,
  {Heber} U.,  2011, \mn@doi [\mnras] {10.1111/j.1365-2966.2011.18786.x}, \href
  {http://adsabs.harvard.edu/abs/2011MNRAS.415.1381C} {415, 1381}

\bibitem[\protect\citeauthoryear{{Deca} et~al.,}{{Deca}
  et~al.}{2012}]{Deca2012}
{Deca} J.,  et~al., 2012, \mn@doi [\mnras] {10.1111/j.1365-2966.2012.20483.x},
  \href {http://adsabs.harvard.edu/abs/2012MNRAS.421.2798D} {421, 2798}

\bibitem[\protect\citeauthoryear{{Deca}, {Vos}, {N{\'e}meth}, {Maxted},
  {Copperwheat}, {Marsh}  \& {{\O}stensen}}{{Deca} et~al.}{2018}]{Deca2018}
{Deca} J.,  {Vos} J.,  {N{\'e}meth} P.,  {Maxted} P.~F.~L.,  {Copperwheat}
  C.~M.,  {Marsh} T.~R.,   {{\O}stensen} R.,  2018, \mn@doi [\mnras]
  {10.1093/mnras/stx2755}, \href
  {http://adsabs.harvard.edu/abs/2018MNRAS.474..433D} {474, 433}

\bibitem[\protect\citeauthoryear{{Dekker}, {D'Odorico}, {Kaufer}, {Delabre}  \&
  {Kotzlowski}}{{Dekker} et~al.}{2000}]{Dekker2000}
{Dekker} H.,  {D'Odorico} S.,  {Kaufer} A.,  {Delabre} B.,   {Kotzlowski} H.,
  2000, in {Iye} M.,  {Moorwood} A.~F.,  eds,  \procspie Vol. 4008, Optical and
  IR Telescope Instrumentation and Detectors. pp 534--545,
  \mn@doi{10.1117/12.395512}

\bibitem[\protect\citeauthoryear{{Foreman-Mackey}, {Hogg}, {Lang}  \&
  {Goodman}}{{Foreman-Mackey} et~al.}{2013}]{Foreman-Mackey2013}
{Foreman-Mackey} D.,  {Hogg} D.~W.,  {Lang} D.,   {Goodman} J.,  2013, \mn@doi
  [\pasp] {10.1086/670067}, \href
  {http://adsabs.harvard.edu/abs/2013PASP..125..306F} {125, 306}

\bibitem[\protect\citeauthoryear{{Freudling}, {Romaniello}, {Bramich},
  {Ballester}, {Forchi}, {Garc{\'{\i}}a-Dabl{\'o}}, {Moehler}  \&
  {Neeser}}{{Freudling} et~al.}{2013}]{reflex2013}
{Freudling} W.,  {Romaniello} M.,  {Bramich} D.~M.,  {Ballester} P.,  {Forchi}
  V.,  {Garc{\'{\i}}a-Dabl{\'o}} C.~E.,  {Moehler} S.,   {Neeser} M.~J.,  2013,
  \mn@doi [\aap] {10.1051/0004-6361/201322494}, \href
  {http://adsabs.harvard.edu/abs/2013A%26A...559A..96F} {559, A96}

\bibitem[\protect\citeauthoryear{{Ge}, {Webbink}, {Chen}  \& {Han}}{{Ge}
  et~al.}{2015}]{Ge2015}
{Ge} H.,  {Webbink} R.~F.,  {Chen} X.,   {Han} Z.,  2015, \mn@doi [ApJ]
  {10.1088/0004-637X/812/1/40}, \href
  {http://adsabs.harvard.edu/abs/2015ApJ...812...40G} {812, 40}

\bibitem[\protect\citeauthoryear{{Geier} et~al.,}{{Geier}
  et~al.}{2011}]{Geier2011}
{Geier} S.,  et~al., 2011, \mn@doi [\aap] {10.1051/0004-6361/201015316}, \href
  {http://adsabs.harvard.edu/abs/2011A%26A...530A..28G} {530, A28}

\bibitem[\protect\citeauthoryear{{Green}, {Liebert}  \& {Saffer}}{{Green}
  et~al.}{2001}]{Green2001}
{Green} E.~M.,  {Liebert} J.,   {Saffer} R.~A.,  2001, in {J.~L.~Provencal,
  H.~L.~Shipman, J.~MacDonald, \& S.~Goodchild} ed.,  ASPCS Vol. 226, 12th
  European Workshop on White Dwarfs. p.~192 (\mn@eprint {}
  {arXiv:astro-ph/0012246})

\bibitem[\protect\citeauthoryear{{Han}, {Tout}  \& {Eggleton}}{{Han}
  et~al.}{2000}]{Han2000}
{Han} Z.,  {Tout} C.~A.,   {Eggleton} P.~P.,  2000, \mn@doi [\mnras]
  {10.1046/j.1365-8711.2000.03839.x}, \href
  {http://adsabs.harvard.edu/abs/2000MNRAS.319..215H} {319, 215}

\bibitem[\protect\citeauthoryear{{Han}, {Podsiadlowski}, {Maxted}, {Marsh}  \&
  {Ivanova}}{{Han} et~al.}{2002}]{Han2002}
{Han} Z.,  {Podsiadlowski} P.,  {Maxted} P.~F.~L.,  {Marsh} T.~R.,   {Ivanova}
  N.,  2002, \mn@doi [\mnras] {10.1046/j.1365-8711.2002.05752.x}, \href
  {http://adsabs.harvard.edu/abs/2002MNRAS.336..449H} {336, 449}

\bibitem[\protect\citeauthoryear{{Han}, {Podsiadlowski}, {Maxted}  \&
  {Marsh}}{{Han} et~al.}{2003}]{Han2003}
{Han} Z.,  {Podsiadlowski} P.,  {Maxted} P.~F.~L.,   {Marsh} T.~R.,  2003,
  \mn@doi [\mnras] {10.1046/j.1365-8711.2003.06451.x}, \href
  {http://adsabs.harvard.edu/abs/2003MNRAS.341..669H} {341, 669}

\bibitem[\protect\citeauthoryear{{Heber}}{{Heber}}{2009}]{Heber2009}
{Heber} U.,  2009, \mn@doi [\araa] {10.1146/annurev-astro-082708-101836}, \href
  {http://adsabs.harvard.edu/abs/2009ARA%26A..47..211H} {47, 211}

\bibitem[\protect\citeauthoryear{{Heber}}{{Heber}}{2016}]{Heber2016}
{Heber} U.,  2016, \mn@doi [\pasp] {10.1088/1538-3873/128/966/082001}, \href
  {http://adsabs.harvard.edu/abs/2016PASP..128h2001H} {128, 082001}

\bibitem[\protect\citeauthoryear{{Hjellming} \& {Webbink}}{{Hjellming} \&
  {Webbink}}{1987}]{Hjellming1987}
{Hjellming} M.~S.,  {Webbink} R.~F.,  1987, \mn@doi [ApJ] {10.1086/165412},
  \href {http://adsabs.harvard.edu/abs/1987ApJ...318..794H} {318, 794}

\bibitem[\protect\citeauthoryear{{Hubeny} \& {Lanz}}{{Hubeny} \&
  {Lanz}}{1995}]{Hubeny1995}
{Hubeny} I.,  {Lanz} T.,  1995, \mn@doi [\apj] {10.1086/175226}, \href
  {http://adsabs.harvard.edu/abs/1995ApJ...439..875H} {439, 875}

\bibitem[\protect\citeauthoryear{{Koen}, {Orosz}  \& {Wade}}{{Koen}
  et~al.}{1998}]{Koen1998}
{Koen} C.,  {Orosz} J.~A.,   {Wade} R.~A.,  1998, \mn@doi [\mnras]
  {10.1046/j.1365-8711.1998.01961.x}, \href
  {http://adsabs.harvard.edu/abs/1998MNRAS.300..695K} {300, 695}

\bibitem[\protect\citeauthoryear{{Kupfer} et~al.,}{{Kupfer}
  et~al.}{2015}]{Kupfer2015}
{Kupfer} T.,  et~al., 2015, VizieR Online Data Catalog, \href
  {http://adsabs.harvard.edu/abs/2015yCat..35760044K} {357}

\bibitem[\protect\citeauthoryear{{Lucy} \& {Sweeney}}{{Lucy} \&
  {Sweeney}}{1971}]{Lucy1971}
{Lucy} L.~B.,  {Sweeney} M.~A.,  1971, \mn@doi [\aj] {10.1086/111159}, \href
  {http://adsabs.harvard.edu/abs/1971AJ.....76..544L} {76, 544}

\bibitem[\protect\citeauthoryear{{Maxted}, {Heber}, {Marsh}  \&
  {North}}{{Maxted} et~al.}{2001}]{Maxted2001}
{Maxted} P.~f.~L.,  {Heber} U.,  {Marsh} T.~R.,   {North} R.~C.,  2001, \mn@doi
  [\mnras] {10.1046/j.1365-8711.2001.04714.x}, \href
  {http://adsabs.harvard.edu/abs/2001MNRAS.326.1391M} {326, 1391}

\bibitem[\protect\citeauthoryear{{Mayor} et~al.,}{{Mayor}
  et~al.}{2003}]{Mayor2003}
{Mayor} M.,  et~al., 2003, The Messenger, \href
  {http://adsabs.harvard.edu/abs/2003Msngr.114...20M} {114, 20}

\bibitem[\protect\citeauthoryear{{Morales-Rueda}, {Maxted}, {Marsh}, {North}
  \& {Heber}}{{Morales-Rueda} et~al.}{2003}]{Morales2003}
{Morales-Rueda} L.,  {Maxted} P.~F.~L.,  {Marsh} T.~R.,  {North} R.~C.,
  {Heber} U.,  2003, \mn@doi [\mnras] {10.1046/j.1365-8711.2003.06088.x}, \href
  {http://adsabs.harvard.edu/abs/2003MNRAS.338..752M} {338, 752}

\bibitem[\protect\citeauthoryear{{Napiwotzki}, {Karl}, {Lisker}, {Heber},
  {Christlieb}, {Reimers}, {Nelemans}  \& {Homeier}}{{Napiwotzki}
  et~al.}{2004}]{Napiwotzki2004}
{Napiwotzki} R.,  {Karl} C.~A.,  {Lisker} T.,  {Heber} U.,  {Christlieb} N.,
  {Reimers} D.,  {Nelemans} G.,   {Homeier} D.,  2004, \mn@doi [\apss]
  {10.1023/B:ASTR.0000044362.07416.6c}, \href
  {http://adsabs.harvard.edu/abs/2004Ap%26SS.291..321N} {291, 321}

\bibitem[\protect\citeauthoryear{{{\O}stensen} \& {Van Winckel}}{{{\O}stensen}
  \& {Van Winckel}}{2012}]{Oestensen2012}
{{\O}stensen} R.~H.,  {Van Winckel} H.,  2012, in {D.~Kilkenny, C.~S.~Jeffery,
  \& C.~Koen} ed.,  ASPCS Vol. 452, Fifth Meeting on Hot Subdwarf Stars and
  Related Objects. p.~163 (\mn@eprint {arXiv} {1112.0977})

\bibitem[\protect\citeauthoryear{{Otani}, {Oswalt}, {Lynas-Gray}, {Kilkenny},
  {Koen}, {Amaral}  \& {Jordan}}{{Otani} et~al.}{2018}]{Otani2018}
{Otani} T.,  {Oswalt} T.~D.,  {Lynas-Gray} A.~E.,  {Kilkenny} D.,  {Koen} C.,
  {Amaral} M.,   {Jordan} R.,  2018, \mn@doi [\apj] {10.3847/1538-4357/aab9bf},
  \href {http://adsabs.harvard.edu/abs/2018ApJ...859..145O} {859, 145}

\bibitem[\protect\citeauthoryear{{Paczynski}}{{Paczynski}}{1976}]{Paczynski1976}
{Paczynski} B.,  1976, in {Eggleton} P.,  {Mitton} S.,   {Whelan} J.,  eds,
  IAU Symposium Vol. 73, Structure and Evolution of Close Binary Systems. p.~75

\bibitem[\protect\citeauthoryear{{Pavlovskii} \& {Ivanova}}{{Pavlovskii} \&
  {Ivanova}}{2015}]{Pavlovskii2015}
{Pavlovskii} K.,  {Ivanova} N.,  2015, \mn@doi [MNRAS] {10.1093/mnras/stv619},
  \href {http://adsabs.harvard.edu/abs/2015MNRAS.449.4415P} {449, 4415}

\bibitem[\protect\citeauthoryear{{Pearson}}{{Pearson}}{1896}]{Pearson1896}
{Pearson} K.,  1896, Philos. Trans. Roy. Soc. London (A), 187, 253

\bibitem[\protect\citeauthoryear{{Tokovinin}, {Fischer}, {Bonati}, {Giguere},
  {Moore}, {Schwab}, {Spronck}  \& {Szymkowiak}}{{Tokovinin}
  et~al.}{2013}]{Tokovinin2013}
{Tokovinin} A.,  {Fischer} D.~A.,  {Bonati} M.,  {Giguere} M.~J.,  {Moore} P.,
  {Schwab} C.,  {Spronck} J.~F.~P.,   {Szymkowiak} A.,  2013, \mn@doi [\pasp]
  {10.1086/674012}, \href {http://adsabs.harvard.edu/abs/2013PASP..125.1336T}
  {125, 1336}

\bibitem[\protect\citeauthoryear{{Vos} et~al.,}{{Vos} et~al.}{2012}]{Vos2012}
{Vos} J.,  et~al., 2012, \mn@doi [\aap] {10.1051/0004-6361/201219723}, \href
  {http://adsabs.harvard.edu/abs/2012A%26A...548A...6V} {548, A6}

\bibitem[\protect\citeauthoryear{{Vos}, {{\O}stensen}, {N{\'e}meth}, {Green},
  {Heber}  \& {Van Winckel}}{{Vos} et~al.}{2013}]{Vos2013}
{Vos} J.,  {{\O}stensen} R.~H.,  {N{\'e}meth} P.,  {Green} E.~M.,  {Heber} U.,
   {Van Winckel} H.,  2013, \mn@doi [\aap] {10.1051/0004-6361/201322200}, \href
  {http://adsabs.harvard.edu/abs/2013A%26A...559A..54V} {559, A54}

\bibitem[\protect\citeauthoryear{{Vos}, {{\O}stensen}, {Vu{\v c}kovi{\'c}}  \&
  {Van Winckel}}{{Vos} et~al.}{2017}]{Vos2017}
{Vos} J.,  {{\O}stensen} R.~H.,  {Vu{\v c}kovi{\'c}} M.,   {Van Winckel} H.,
  2017, \mn@doi [\aap] {10.1051/0004-6361/201730958}, \href
  {http://adsabs.harvard.edu/abs/2017A%26A...605A.109V} {605, A109}

\bibitem[\protect\citeauthoryear{{Vos}, {N{\'e}meth}, {Vuc{\v c}kovi{\'c}},
  {{\O}stensen}  \& {Parsons}}{{Vos} et~al.}{2018}]{Vos2018a}
{Vos} J.,  {N{\'e}meth} P.,  {Vuc{\v c}kovi{\'c}} M.,  {{\O}stensen} R.,
  {Parsons} S.,  2018, \mn@doi [\mnras] {10.1093/mnras/stx2198}, \href
  {http://adsabs.harvard.edu/abs/2018MNRAS.473..693V} {473, 693}

\bibitem[\protect\citeauthoryear{{Webbink}}{{Webbink}}{1984}]{Webbink1984}
{Webbink} R.~F.,  1984, \mn@doi [\apj] {10.1086/161701}, \href
  {http://adsabs.harvard.edu/abs/1984ApJ...277..355W} {277, 355}

\bibitem[\protect\citeauthoryear{{Woods}, {Ivanova}, {van der Sluys}  \&
  {Chaichenets}}{{Woods} et~al.}{2012}]{Woods2012}
{Woods} T.~E.,  {Ivanova} N.,  {van der Sluys} M.~V.,   {Chaichenets} S.,
  2012, \mn@doi [ApJ] {10.1088/0004-637X/744/1/12}, \href
  {http://adsabs.harvard.edu/abs/2012ApJ...744...12W} {744, 12}

\makeatother
\end{thebibliography}

%%%%%%%%%%%%%%%%%%%%%%%%%%%%%%%%%%%%%%%%%%%%%%%%%%

%%%%%%%%%%%%%%%%%%%%% APPENDIX %%%%%%%%%%%%%%%%%%%

\appendix

\section{Radial velocity data}

\begin{table}
   \centering
   \caption{Radial velocities of PB\,6355} \label{tb:rv_PB6355}
   \begin{tabular}{lr@{ $\pm$ }lr@{ $\pm$ }ll}
    \hline
    BJD      & \multicolumn{2}{c}{MS}            &  \multicolumn{2}{c}{sdB}           &  Instrument  \\
    -2450000 & \multicolumn{2}{c}{(km s$^{-1}$)} &  \multicolumn{2}{c}{(km s$^{-1}$)} &              \\\hline\hline
    6989.5763  &  2.51  &  0.73  &  -15.62  &  1.90  &  feros  \\ 
    6996.6091  &  4.08  &  0.82  &  -9.43  &  2.10  &  feros  \\ 
    7308.7335  &  -5.97  &  0.37  &  17.84  &  0.82  &  feros  \\ 
    7312.6271  &  -6.21  &  0.43  &  18.64  &  0.76  &  feros  \\ 
    7599.8215  &  3.58  &  0.67  &  -8.07  &  1.60  &  feros  \\ 
    7619.7301  &  2.33  &  0.30  &  -11.23  &  0.72  &  uves  \\ 
    7663.6286  &  2.86  &  0.22  &  -10.87  &  0.73  &  uves  \\ 
    7725.5654  &  2.06  &  0.42  &  -10.15  &  0.91  &  uves  \\ 
    7919.9058  &  -3.06  &  0.28  &  8.16  &  0.81  &  uves  \\ 
    7969.8943  &  -5.37  &  0.28  &  14.12  &  0.81  &  uves  \\ 
    8013.7136  &  -7.65  &  0.32  &  18.95  &  0.71  &  uves  \\ 
    \hline
   \end{tabular}
\end{table}

\begin{table}
   \centering
   \caption{Radial velocities of MCT\,0146--2651} \label{tb:rv_MCT0146-2651}
   \begin{tabular}{lr@{ $\pm$ }lr@{ $\pm$ }ll}
    \hline
    BJD      & \multicolumn{2}{c}{MS}            &  \multicolumn{2}{c}{sdB}           &  Instrument  \\
    -2450000 & \multicolumn{2}{c}{(km s$^{-1}$)} &  \multicolumn{2}{c}{(km s$^{-1}$)} &              \\\hline\hline
    5846.6217  &  33.74  &  0.15  &  51.17  &  0.37  &  uves  \\ 
    5908.5755  &  34.54  &  0.24  &  48.42  &  0.26  &  uves  \\ 
    5934.5897  &  35.83  &  0.12  &  47.44  &  0.23  &  uves  \\ 
    6842.9088  &  43.56  &  0.17  &  36.02  &  0.28  &  uves  \\ 
    7034.5375  &  45.41  &  0.24  &  34.78  &  2.27  &  chiron  \\ 
    7051.5847  &  44.76  &  0.26  &  36.81  &  2.67  &  chiron  \\ 
    7210.8916  &  36.52  &  0.38  &  45.35  &  2.47  &  chiron  \\ 
    7361.6492  &  33.90  &  0.24  &  49.65  &  2.44  &  chiron  \\ 
    7379.5851  &  33.86  &  0.24  &  51.83  &  1.34  &  chiron  \\ 
    7403.5678  &  34.31  &  0.22  &  51.08  &  2.28  &  chiron  \\ 
    7938.8378  &  39.12  &  0.15  & \multicolumn{2}{c}{/} &  feros  \\ 
    8004.8137  &  36.28  &  0.13  & \multicolumn{2}{c}{/} &  feros  \\ 
    8134.5376  &  34.14  &  0.36  &  51.80  &  2.10  &  chiron  \\ 
    \hline
   \end{tabular}
\end{table}

\begin{table}
   \centering
   \caption{Radial velocities of FAUST\,321} \label{tb:rv_J01513-7548}
   \begin{tabular}{lr@{ $\pm$ }lr@{ $\pm$ }ll}
    \hline
    BJD      & \multicolumn{2}{c}{MS}            &  \multicolumn{2}{c}{sdB}           &  Instrument  \\
    -2450000 & \multicolumn{2}{c}{(km s$^{-1}$)} &  \multicolumn{2}{c}{(km s$^{-1}$)} &              \\\hline\hline
    5846.6283  &  -43.94  &  0.21  &  -23.96  &  0.85  &  uves  \\ 
    5849.6214  &  -43.82  &  0.24  &  -21.87  &  1.00  &  uves  \\ 
    6842.9173  &  -44.13  &  0.15  &  -25.23  &  0.77  &  uves  \\ 
    7328.5897  &  -31.90  &  0.17  &  -48.53  &  0.56  &  uves  \\ 
    7435.5526  &  -33.21  &  0.17  &  -50.32  &  0.51  &  uves  \\ 
    7621.8357  &  -40.23  &  0.14  &  -34.78  &  0.29  &  uves  \\ 
    7665.5538  &  -41.67  &  0.18  &  -28.02  &  0.50  &  uves  \\ 
    7725.7164  &  -42.92  &  0.13  &  -28.29  &  0.42  &  uves  \\ 
    7821.5242  &  -43.80  &  0.18  &  -23.53  &  0.66  &  uves  \\ 
    7921.8876  &  -43.32  &  0.18  &  -26.60  &  0.54  &  uves  \\ 
    7938.8714  &  -43.20  &  0.15  &  -26.07  &  0.57  &  uves  \\ 
    7968.8977  &  -42.50  &  0.20  &  -27.63  &  0.84  &  uves  \\ 
    8003.8379  &  -41.88  &  0.28  &  -29.75  &  0.92  &  uves  \\ 
    \hline
   \end{tabular}
\end{table}

\begin{table}
   \centering
   \caption{Radial velocities of JL\,277} \label{tb:rv_JL277}
   \begin{tabular}{lr@{ $\pm$ }lr@{ $\pm$ }ll}
    \hline
    BJD      & \multicolumn{2}{c}{MS}            &  \multicolumn{2}{c}{sdB}           &  Instrument  \\
    -2450000 & \multicolumn{2}{c}{(km s$^{-1}$)} &  \multicolumn{2}{c}{(km s$^{-1}$)} &              \\\hline\hline
    5849.6371  &  108.82  &  0.38  &  88.46  &  0.45  &  uves  \\ 
    5934.6082  &  106.15  &  0.40  &  95.25  &  0.45  &  uves  \\ 
    6848.8772  &  109.59  &  0.26  &  87.44  &  0.25  &  uves  \\ 
    7328.5593  &  96.97  &  0.33  &  118.24  &  0.43  &  uves  \\ 
    7382.6439  &  96.71  &  0.23  &  117.82  &  0.44  &  uves  \\ 
    7435.5407  &  97.01  &  0.35  &  117.25  &  0.56  &  uves  \\ 
    7595.8753  &  99.84  &  0.37  &  110.58  &  0.41  &  uves  \\ 
    7621.7928  &  100.87  &  0.30  &  108.98  &  0.29  &  uves  \\ 
    7643.6870  &  100.70  &  0.23  &  106.91  &  0.30  &  uves  \\ 
    7665.5653  &  101.23  &  0.32  &  105.37  &  0.50  &  uves  \\ 
    7674.7067  &  102.42  &  0.32  &  105.39  &  0.43  &  uves  \\ 
    7712.5963  &  103.28  &  0.36  &  102.21  &  0.36  &  uves  \\ 
    7807.5569  &  106.74  &  0.50  &  94.54  &  0.73  &  uves  \\ 
    7893.9116  &  109.00  &  0.36  &  89.24  &  0.38  &  uves  \\ 
    7938.8416  &  109.84  &  0.19  &  86.97  &  0.48  &  uves  \\ 
    7969.7878  &  109.83  &  0.26  &  87.56  &  0.33  &  uves  \\ 
    \hline
   \end{tabular}
\end{table}

\begin{table}
   \centering
   \caption{Radial velocities of J\,02286--3625} \label{tb:rv_J02286-3625}
   \begin{tabular}{lr@{ $\pm$ }lr@{ $\pm$ }ll}
    \hline
    BJD      & \multicolumn{2}{c}{MS}            &  \multicolumn{2}{c}{sdB}           &  Instrument  \\
    -2450000 & \multicolumn{2}{c}{(km s$^{-1}$)} &  \multicolumn{2}{c}{(km s$^{-1}$)} &              \\\hline\hline
    6841.9031  &  -6.13  &  2.57  &  -4.33  &  0.33  &  uves  \\ 
    6873.8016  &  -5.06  &  1.44  &  -9.47  &  0.36  &  uves  \\ 
    6903.8547  &  -3.93  &  1.14  &  -13.78  &  0.30  &  uves  \\ 
    7328.5458  &  -10.54  &  0.95  &  5.77  &  0.60  &  uves  \\ 
    7382.6308  &  -8.62  &  1.26  &  -3.44  &  0.54  &  uves  \\ 
    7435.5264  &  -3.82  &  0.86  &  -10.61  &  0.60  &  uves  \\ 
    7587.7744  &  -1.28  &  2.05  &  -17.07  &  0.51  &  uves  \\ 
    7621.7006  &  -0.40  &  0.72  &  -12.57  &  0.45  &  uves  \\ 
    7643.8083  &  -3.14  &  0.99  &  -7.17  &  0.43  &  uves  \\ 
    7680.8533  &  -11.46  &  1.07  &  2.59  &  0.42  &  uves  \\ 
    7725.5466  &  -15.21  &  1.39  &  13.44  &  0.41  &  uves  \\ 
    7821.5387  &  -15.38  &  2.01  &  13.11  &  0.92  &  uves  \\ 
    7919.8925  &  -5.77  &  2.13  &  0.56  &  0.64  &  uves  \\ 
    7948.8644  &  -5.00  &  1.10  &  -4.69  &  0.47  &  uves  \\ 
    7976.7658  &  -4.08  &  1.66  &  -8.86  &  0.67  &  uves  \\ 
    7981.7126  &  -5.14  &  1.99  &  -9.16  &  0.38  &  uves  \\ 
    \hline
   \end{tabular}
\end{table}

\begin{table}
   \centering
   \caption{Radial velocities of EC\,03143--5945} \label{tb:rv_EC03143-5945}
   \begin{tabular}{lr@{ $\pm$ }lr@{ $\pm$ }ll}
    \hline
    BJD      & \multicolumn{2}{c}{MS}            &  \multicolumn{2}{c}{sdB}           &  Instrument  \\
    -2450000 & \multicolumn{2}{c}{(km s$^{-1}$)} &  \multicolumn{2}{c}{(km s$^{-1}$)} &              \\\hline\hline
    3953.8771  &  38.59  &  0.50  &  45.71  &  1.00  &  feros  \\ 
    5500.7301  &  41.38  &  0.50  &  42.25  &  1.00  &  feros  \\ 
    6840.8894  &  33.15  &  0.40  &  58.38  &  0.29  &  uves  \\ 
    6871.8719  &  33.10  &  0.24  &  58.44  &  0.31  &  uves  \\ 
    6901.7816  &  33.61  &  0.60  &  57.68  &  0.24  &  uves  \\ 
    7328.6004  &  47.33  &  0.48  &  25.13  &  0.39  &  uves  \\ 
    7382.6563  &  46.29  &  0.58  &  24.97  &  0.22  &  uves  \\ 
    7435.5636  &  45.20  &  0.72  &  28.70  &  0.50  &  uves  \\ 
    7611.8867  &  39.15  &  0.48  &  42.95  &  0.26  &  uves  \\ 
    7639.6391  &  38.28  &  0.44  &  45.06  &  0.32  &  uves  \\ 
    7723.5799  &  35.80  &  0.36  &  52.13  &  0.21  &  uves  \\ 
    7807.5708  &  33.75  &  0.44  &  56.64  &  0.39  &  uves  \\ 
    7831.5241  &  33.87  &  0.48  &  57.66  &  0.36  &  uves  \\ 
    7850.4960  &  33.88  &  0.52  &  58.28  &  0.39  &  uves  \\ 
    7938.8556  &  33.37  &  0.40  &  56.73  &  0.42  &  uves  \\ 
    7973.7688  &  34.00  &  0.52  &  55.78  &  0.28  &  uves  \\ 
    \hline
   \end{tabular}
\end{table}

\begin{table}
   \centering
   \caption{Radial velocities of J\,03322--0233} \label{tb:rv_J03322-0233}
   \begin{tabular}{lr@{ $\pm$ }lr@{ $\pm$ }ll}
    \hline
    BJD      & \multicolumn{2}{c}{MS}            &  \multicolumn{2}{c}{sdB}           &  Instrument  \\
    -2450000 & \multicolumn{2}{c}{(km s$^{-1}$)} &  \multicolumn{2}{c}{(km s$^{-1}$)} &              \\\hline\hline
    6242.6946  &  23.08  &  0.75  &  28.09  &  0.36  &  feros  \\ 
    6244.6982  &  21.58  &  0.75  &  27.37  &  0.36  &  feros  \\ 
    7340.6774  &  18.84  &  0.57  &  45.27  &  0.38  &  uves  \\ 
    7382.6168  &  18.95  &  0.73  &  40.50  &  0.60  &  uves  \\ 
    7436.5564  &  22.92  &  1.11  &  34.41  &  0.64  &  uves  \\ 
    7619.8837  &  27.45  &  0.52  &  17.07  &  0.68  &  uves  \\ 
    7941.9136  &  28.48  &  0.72  &  17.70  &  0.53  &  uves  \\ 
    7986.8956  &  27.97  &  0.46  &  20.02  &  0.47  &  uves  \\ 
    8017.8291  &  27.70  &  0.60  &  21.27  &  0.44  &  uves  \\ 
    \hline
   \end{tabular}
\end{table}

\begin{table}
   \centering
   \caption{Radial velocities of J\,05396--2833} \label{tb:rv_J05396-2833}
   \begin{tabular}{lr@{ $\pm$ }lr@{ $\pm$ }ll}
    \hline
    BJD      & \multicolumn{2}{c}{MS}            &  \multicolumn{2}{c}{sdB}           &  Instrument  \\
    -2450000 & \multicolumn{2}{c}{(km s$^{-1}$)} &  \multicolumn{2}{c}{(km s$^{-1}$)} &              \\\hline\hline
    5849.7885  &  8.85  &  0.25  &  26.72  &  0.73  &  uves  \\ 
    5934.6182  &  11.60  &  0.22  &  18.57  &  0.26  &  uves  \\ 
    7756.6487  &  17.53  &  0.19  &  13.89  &  0.43  &  uves  \\ 
    7798.6864  &  19.50  &  0.26  &  11.09  &  0.87  &  uves  \\ 
    7846.5680  &  20.91  &  0.21  &  5.86  &  0.43  &  uves  \\ 
    7989.8761  &  20.61  &  0.17  &  10.08  &  0.50  &  uves  \\ 
    8022.7543  &  18.73  &  0.13  &  10.06  &  0.40  &  uves  \\ 
    \hline
   \end{tabular}
\end{table}

\begin{table}
   \centering
   \caption{Radial velocities of PG\,1514+034} \label{tb:rv_PG1514+034}
   \begin{tabular}{lr@{ $\pm$ }lr@{ $\pm$ }ll}
    \hline
    BJD      & \multicolumn{2}{c}{MS}            &  \multicolumn{2}{c}{sdB}           &  Instrument  \\
    -2450000 & \multicolumn{2}{c}{(km s$^{-1}$)} &  \multicolumn{2}{c}{(km s$^{-1}$)} &              \\\hline\hline
    6759.7200  &  -63.25  &  0.18  &  -86.00  &  0.36  &  uves  \\ 
    6803.7327  &  -69.83  &  0.40  &  -72.76  &  0.98  &  uves  \\ 
    6814.4957  &  -71.10  &  0.20  &  -73.35  &  0.42  &  uves  \\ 
    6870.5995  &  -77.57  &  0.20  &  -61.03  &  0.50  &  uves  \\ 
    7439.8587  &  -81.45  &  0.34  &  -54.73  &  0.48  &  uves  \\ 
    7797.7992  &  -74.05  &  0.24  &  -67.10  &  1.16  &  uves  \\ 
    7831.8187  &  -77.80  &  0.14  &  -60.11  &  0.40  &  uves  \\ 
    7864.8116  &  -80.65  &  0.22  &  -56.73  &  0.28  &  uves  \\ 
    7913.6756  &  -81.36  &  0.26  &  -53.30  &  0.46  &  uves  \\ 
    7947.5698  &  -79.95  &  0.22  &  -56.22  &  0.48  &  uves  \\ 
    8168.3540  &  -60.53  &  0.30  &  -89.10  &  0.30  &  uves  \\ 
    \hline
   \end{tabular}
\end{table}

\begin{table}
   \centering
   \caption{Radial velocities of J\,16287+1118} \label{tb:rv_J16287+1118}
   \begin{tabular}{lr@{ $\pm$ }lr@{ $\pm$ }ll}
    \hline
    BJD      & \multicolumn{2}{c}{MS}            &  \multicolumn{2}{c}{sdB}           &  Instrument  \\
    -2450000 & \multicolumn{2}{c}{(km s$^{-1}$)} &  \multicolumn{2}{c}{(km s$^{-1}$)} &              \\\hline\hline
    6505.5615  &  -41.05  &  0.18  & \multicolumn{2}{c}{/} &  feros  \\ 
    6509.5547  &  -41.38  &  0.14  & \multicolumn{2}{c}{/} &  feros  \\ 
    7440.8701  &  -40.23  &  0.32  &  -47.52  &  0.98  &  uves  \\ 
    7455.8771  &  -39.80  &  0.16  &  -47.84  &  0.65  &  uves  \\ 
    7799.8470  &  -43.97  &  0.28  &  -40.30  &  1.24  &  uves  \\ 
    7894.8135  &  -45.72  &  0.18  &  -33.46  &  0.66  &  uves  \\ 
    7913.7705  &  -46.12  &  0.30  &  -34.76  &  1.22  &  uves  \\ 
    7947.6061  &  -46.47  &  0.14  &  -31.60  &  0.54  &  uves  \\ 
    7982.5465  &  -46.51  &  0.14  &  -32.70  &  0.55  &  uves  \\ 
    \hline
   \end{tabular}
\end{table}

\begin{table}
   \centering
   \caption{Radial velocities of PG\,2148+095} \label{tb:rv_PG2148+095}
   \begin{tabular}{lr@{ $\pm$ }lr@{ $\pm$ }ll}
    \hline
    BJD      & \multicolumn{2}{c}{MS}            &  \multicolumn{2}{c}{sdB}           &  Instrument  \\
    -2450000 & \multicolumn{2}{c}{(km s$^{-1}$)} &  \multicolumn{2}{c}{(km s$^{-1}$)} &              \\\hline\hline
    4015.5894  &  -137.99  &  0.36  & \multicolumn{2}{c}{/} &  feros  \\ 
    4037.6033  &  -136.75  &  0.41  & \multicolumn{2}{c}{/} &  feros  \\ 
    6841.8918  &  -137.99  &  0.28  &  -150.11  &  1.18  &  uves  \\ 
    6873.7610  &  -136.55  &  0.26  &  -152.46  &  1.73  &  uves  \\ 
    6906.6847  &  -136.95  &  0.36  &  -151.42  &  2.05  &  uves  \\ 
    7340.5185  &  -141.44  &  0.52  &  -141.12  &  2.84  &  uves  \\ 
    7611.8380  &  -147.96  &  0.38  &  -118.69  &  2.12  &  uves  \\ 
    7643.6486  &  -148.31  &  0.22  &  -118.12  &  1.67  &  uves  \\ 
    7901.8972  &  -143.50  &  0.22  &  -134.13  &  1.67  &  uves  \\ 
    7941.6897  &  -142.74  &  0.28  &  -135.04  &  1.62  &  uves  \\ 
    7981.7270  &  -141.83  &  0.22  &  -136.85  &  1.22  &  uves  \\ 
    \hline
   \end{tabular}
\end{table}

\newpage

\begin{figure}
    \includegraphics{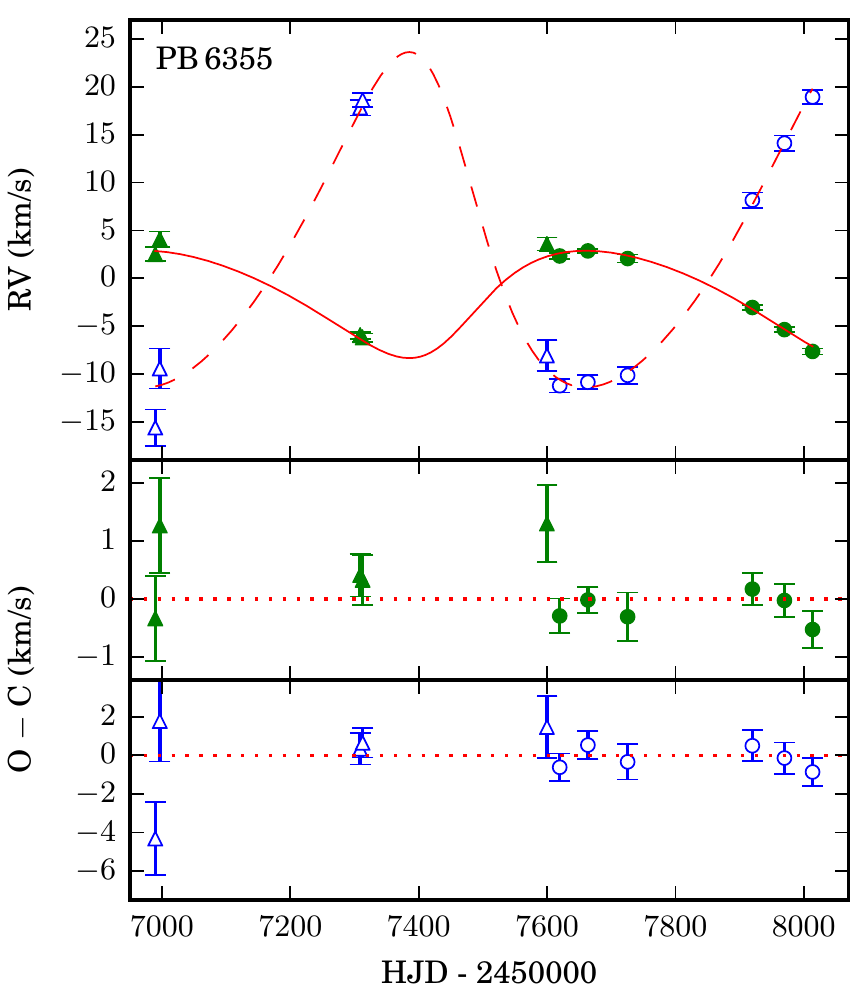}
    \caption{
The radial velocity curves and residuals (O$-$C) for PB\,6355. The radial velocities of the cool companion are plotted in green filled symbols, while those of the sdB are shown in open blue symbols. The best fitting Keplerian orbit is shown in red full line for the cool companion and red dashed line for the sdB. Radial velocities derived from UVES spectra are shown as circles, those of FEROS spectra are shown as triangles and those from the CHIRON spectra in squares.
}
    \label{fig:rv_PB6355}
\end{figure}

\begin{figure}
    \includegraphics{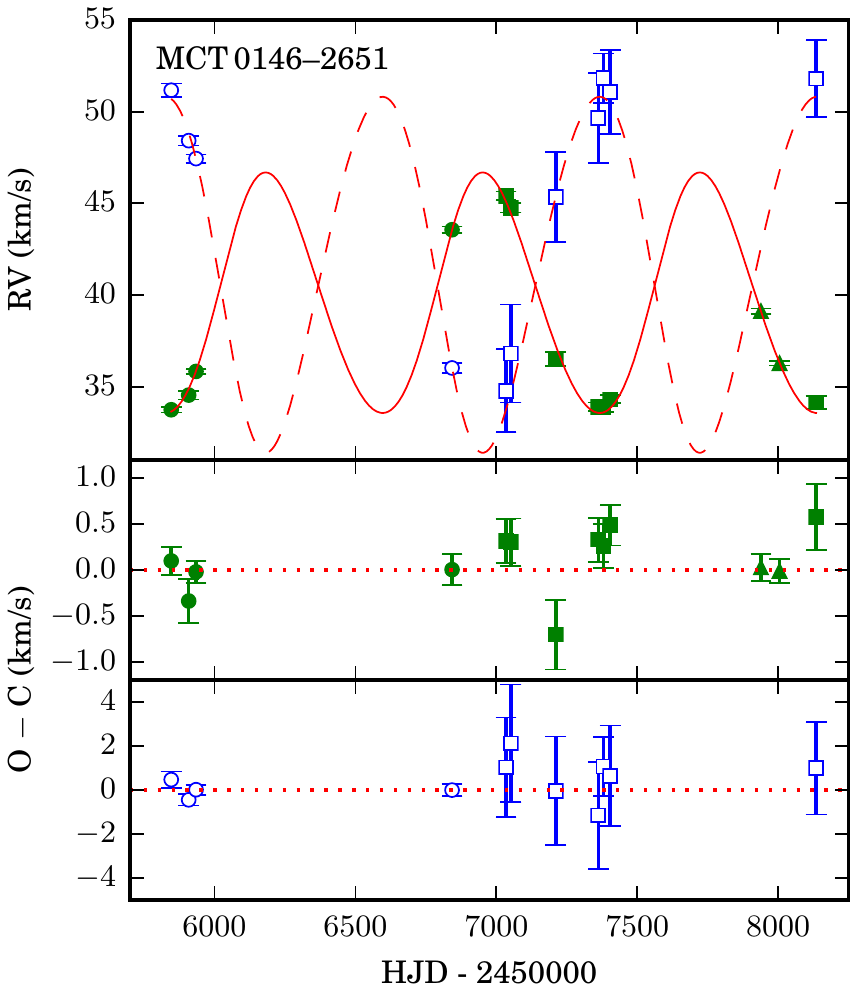}
    \caption{RV curves for MCT\,0146--2651, same as Fig.\,\ref{fig:rv_PB6355}}
    \label{fig:rv_MCT0146-2651}
\end{figure}

\begin{figure}
    \includegraphics{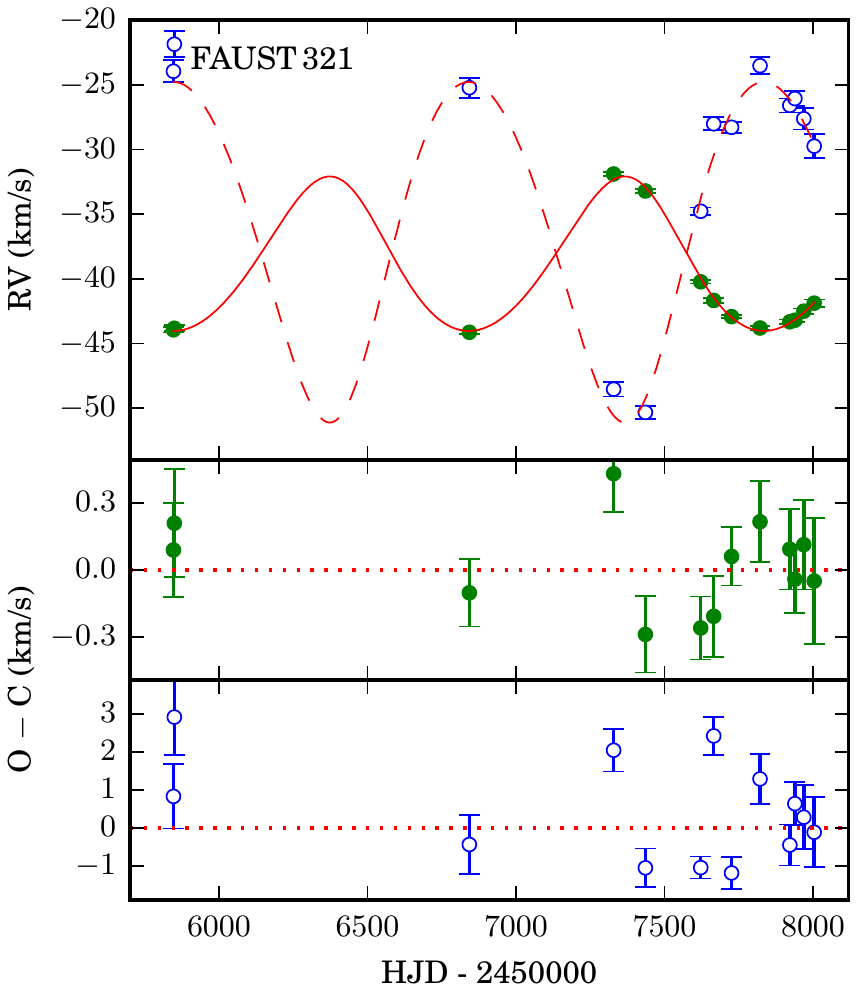}
    \caption{RV curves for FAUST\,321, same as Fig.\,\ref{fig:rv_PB6355}}
    \label{fig:rv_J01513-7548}
\end{figure}

\begin{figure}
    \includegraphics{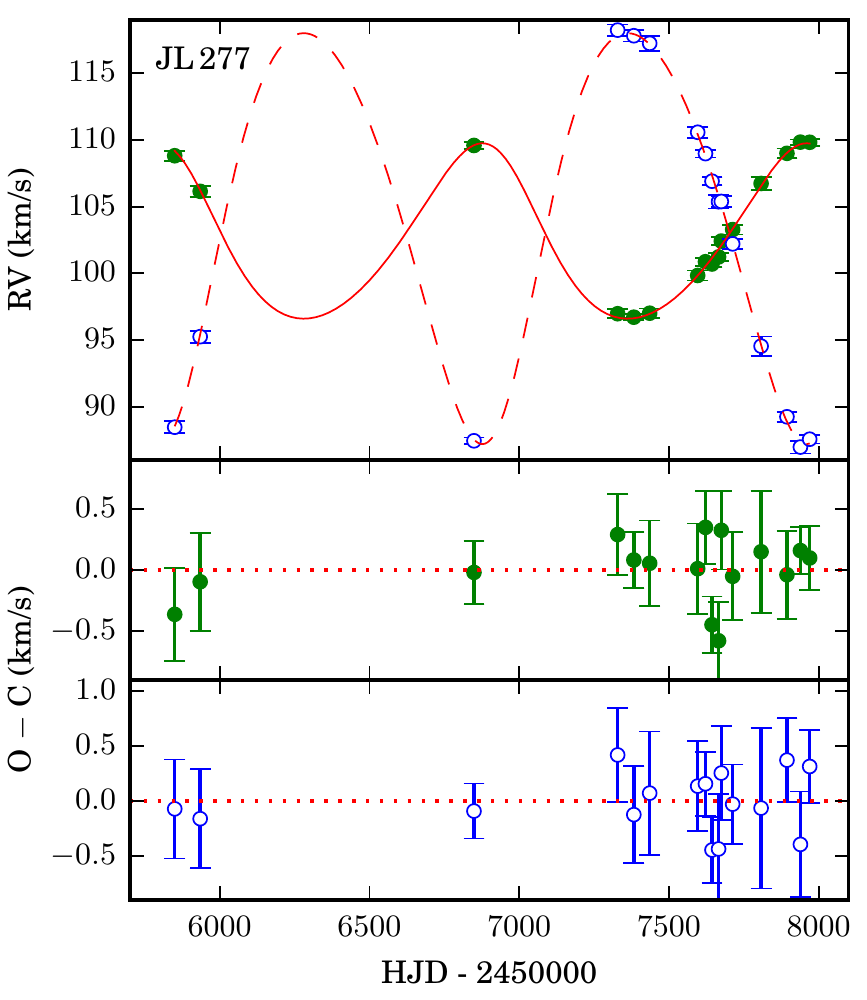}
    \caption{RV curves for JL\,277, same as Fig.\,\ref{fig:rv_PB6355}}
    \label{fig:rv_JL277}
\end{figure}

\begin{figure}
    \includegraphics{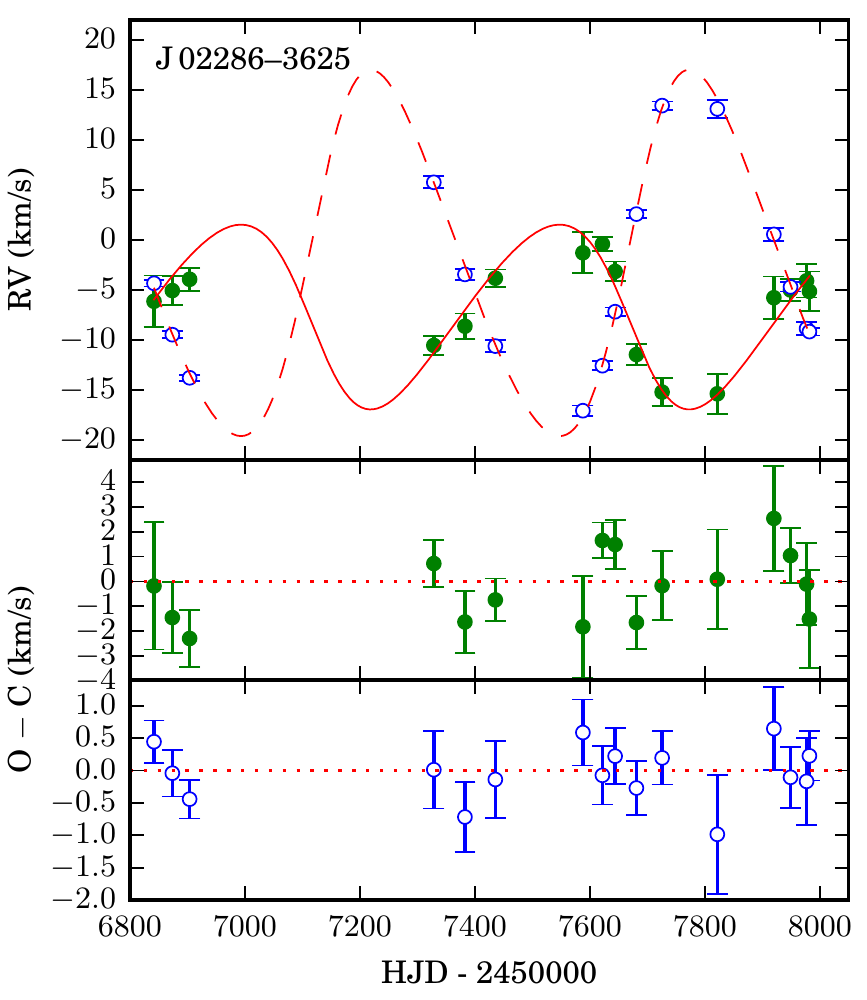}
    \caption{RV curves for J\,02286--3625, same as Fig.\,\ref{fig:rv_PB6355}}
    \label{fig:rv_J02286-3625}
\end{figure}

\begin{figure}
    \includegraphics{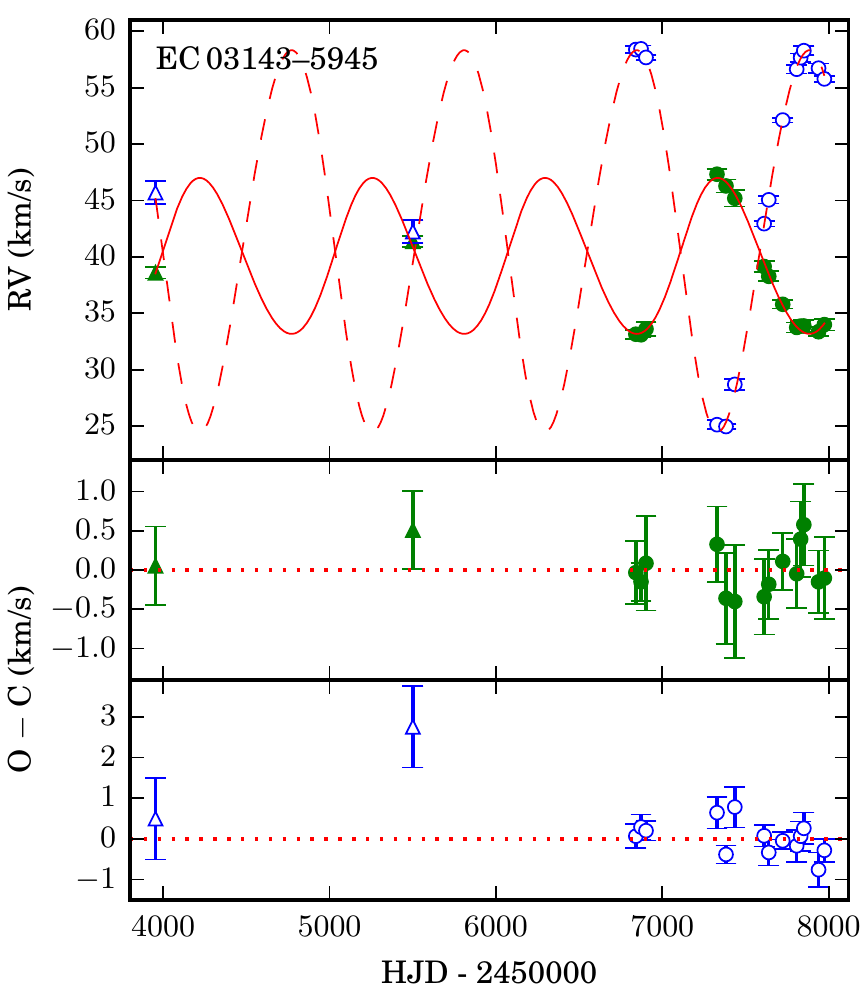}
    \caption{RV curves for EC\,03143--5945, same as Fig.\,\ref{fig:rv_PB6355}}
    \label{fig:rv_EC03143-5945}
\end{figure}

\begin{figure}
    \includegraphics{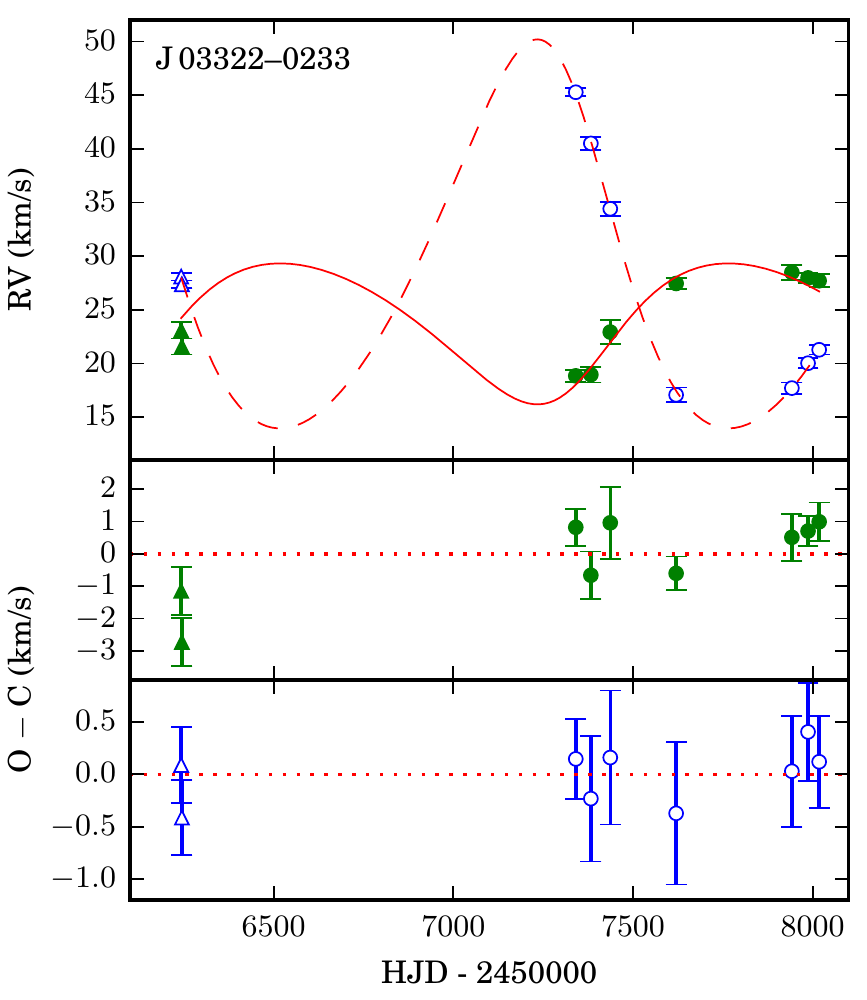}
    \caption{RV curves for J\,03322--0233, same as Fig.\,\ref{fig:rv_PB6355}}
    \label{fig:rv_J03322-0233}
\end{figure}

\begin{figure}
    \includegraphics{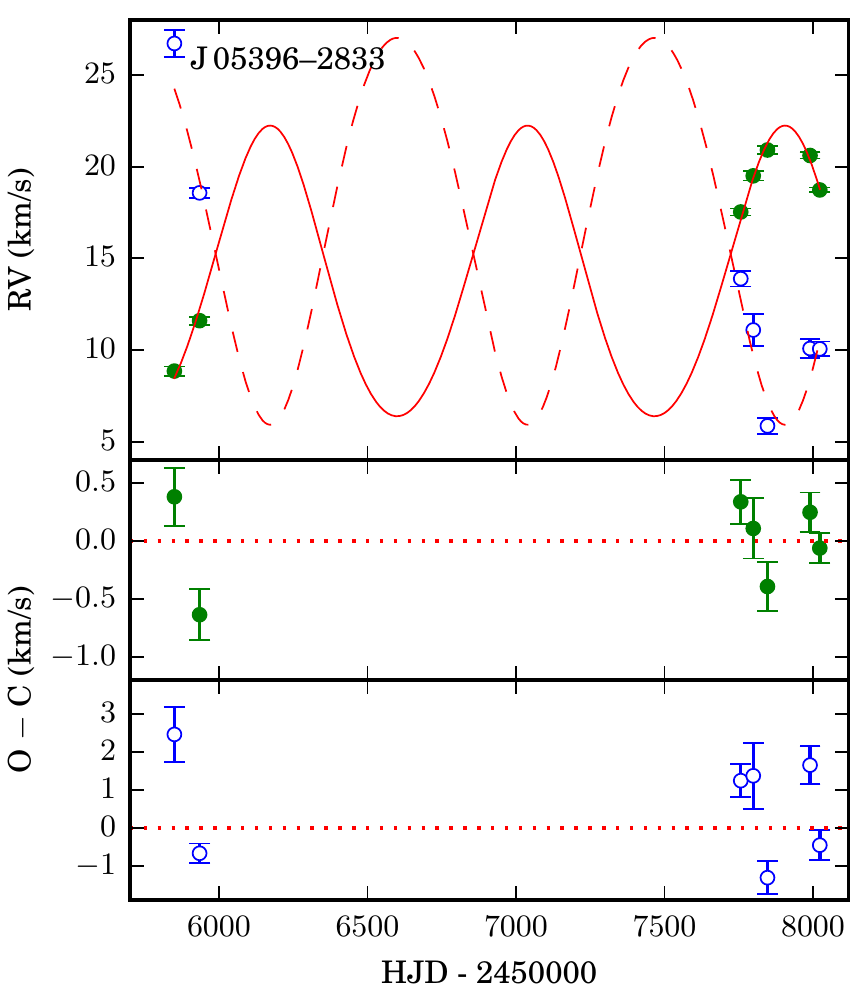}
    \caption{RV curves for J\,05396--2833, same as Fig.\,\ref{fig:rv_PB6355}}
    \label{fig:rv_J05396-2833}
\end{figure}

\begin{figure}
    \includegraphics{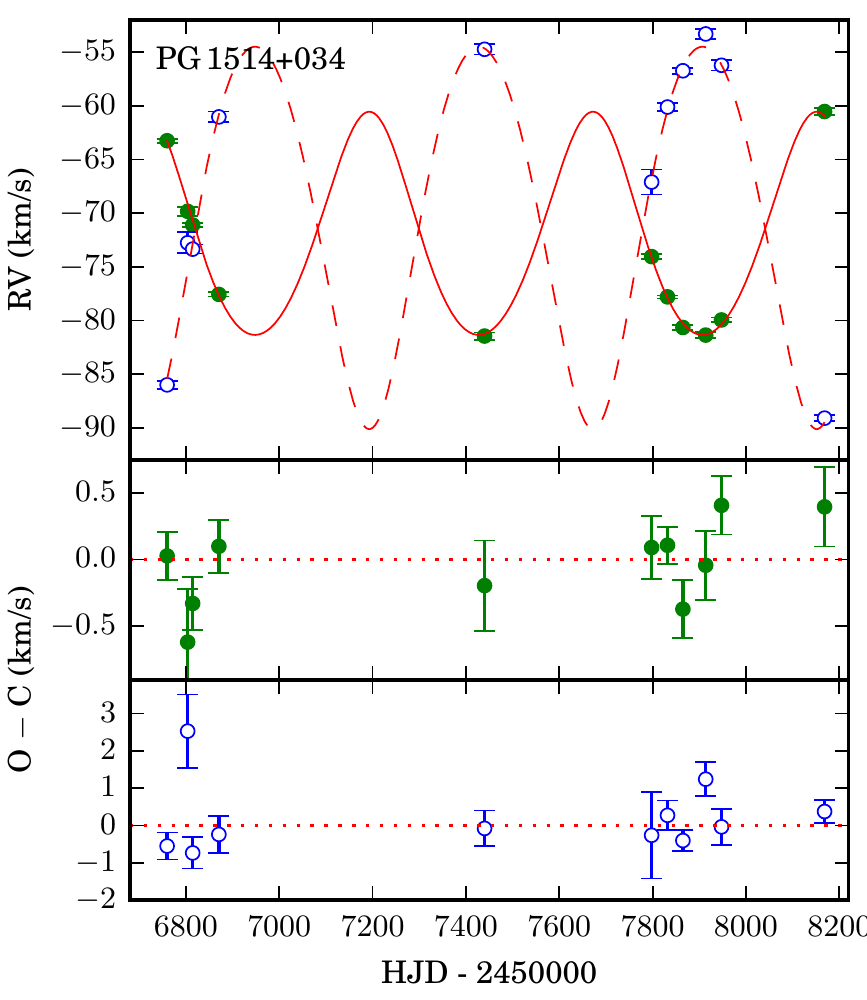}
    \caption{RV curves for PG\,1514+034, same as Fig.\,\ref{fig:rv_PB6355}}
    \label{fig:rv_PG1514+034}
\end{figure}

\begin{figure}
    \includegraphics{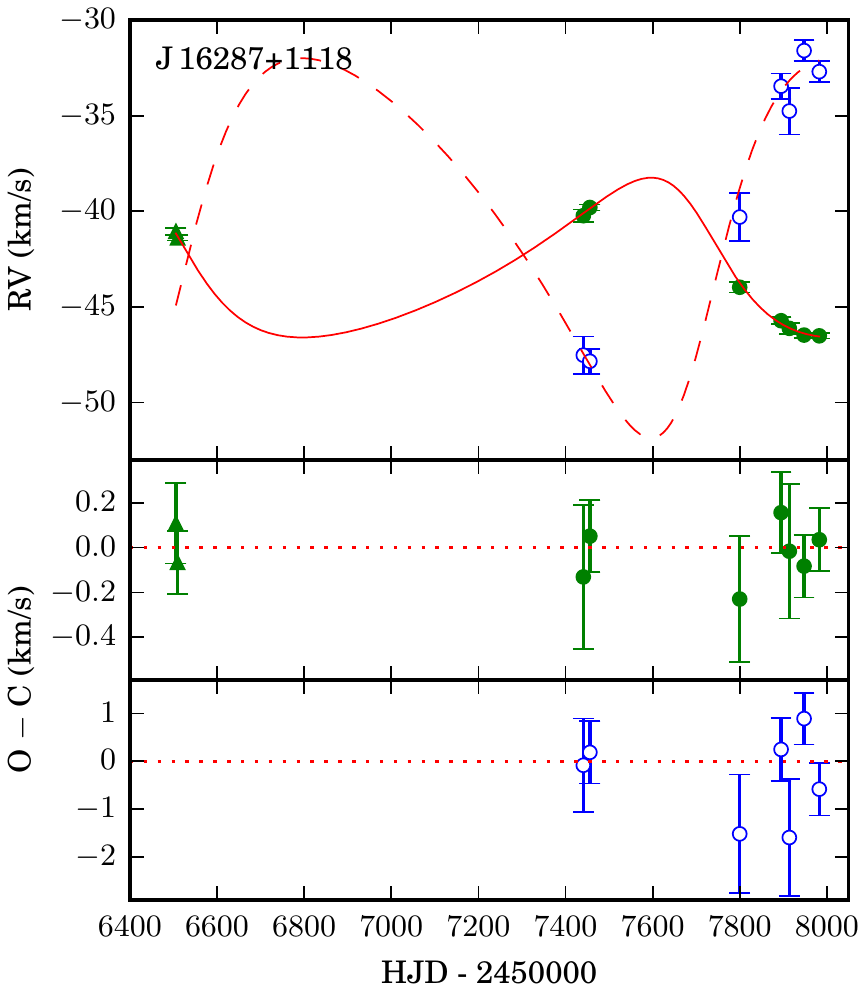}
    \caption{RV curves for J\,16287+1118, same as Fig.\,\ref{fig:rv_PB6355}}
    \label{fig:rv_J16287+1118}
\end{figure}

\begin{figure}
    \includegraphics{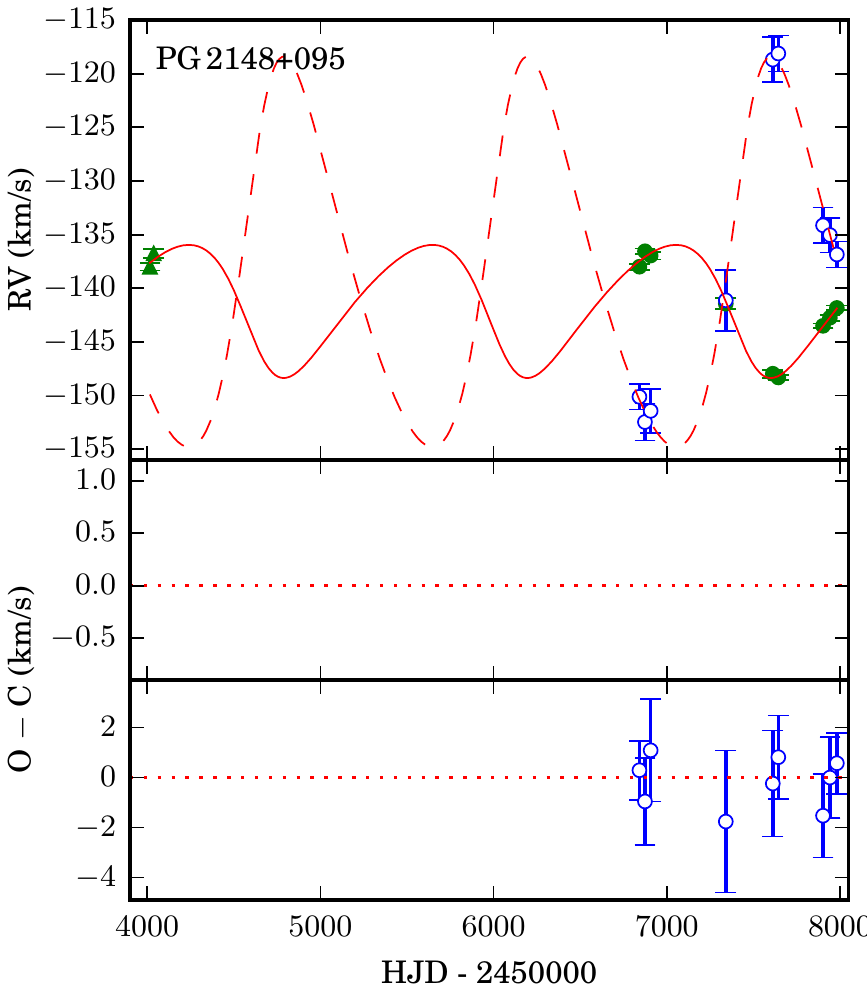}
    \caption{RV curves for PG\,2148+095, same as Fig.\,\ref{fig:rv_PB6355}}
    \label{fig:rv_PG2148+095}
\end{figure}

% Don't change these lines
\bsp	% typesetting comment
\label{lastpage}
\end{document}